\documentstyle[floats,prd,aps,epsf]{revtex}

\draft 
\begin{document}
\def\go{\mathrel{\raise.4ex\hbox{$>$}\mkern-14mu\lower0.7ex\hbox{$\sim$}}}
\def\lo{\mathrel{\raise.4ex\hbox{$<$}\mkern-14mu\lower0.7ex\hbox{$\sim$}}}
\newcommand{\ns}{{\mathit NS}}
\newcommand{\bh}{{\mathit BH}}
\newcommand{\gr}{{\mathit GR}}
\newcommand{\inin}{{\mathrm in}}
\newcommand{\ouou}{{\mathrm out}}
\newcommand{\Jseq}{J(d_G)}
\newcommand{\bJseq}{\bar{J}(\bar{d}_G)}
\newcommand{\dg}{d_G}
\newcommand{\dse}{d_{{\rm sec}}}
\newcommand{\ddy}{d_{{\rm dyn}}}
\newcommand{\dR}{d_R}
\newcommand{\whd}{\widehat{d}}
\newcommand{\bdg}{\bar{d}_G}
\newcommand{\bdse}{\bar{d}_{{\rm sec}}}
\newcommand{\bddy}{\bar{d}_{{\rm dyn}}}
\newcommand{\bdR}{\bar{d}_R}
\newcommand{\bM}{\bar{M}}
\newcommand{\bJ}{\bar{J}}
\newcommand{\bMo}{\bar{M}_0}
\newcommand{\bMotot}{\bar{M}_{0,\rm tot}}
\newcommand{\Motot}{{M}_{0,\rm tot}}
\newcommand{\madm}{M_{\rm ADM}}
\newcommand{\rmfrac}{M_0(j)/M_0}
\newcommand{\jisco}{j_{\rm ISCO}}
\newcommand{\eee}{(e_{\rm max} - e_\infty)/e_\infty}
\newcommand{\brmx}{\bar{\rho}_{\rm max}}
\newcommand{\compa}{(M/R)_\infty} 
\newcommand{\compala}{\left(\frac{M}{R}\right)_\infty} 
\newcommand{\jom}{J_{\rm tot}/M_{\rm ADM}^2}
\newcommand{\fgw}{f_{{\rm GW}}}
\newcommand{\beq}{\begin{equation}}
\newcommand{\eeq}{\end{equation}}
\newcommand{\beqn}{\begin{eqnarray}}
\newcommand{\eeqn}{\end{eqnarray}}
\newcommand{\pa}{\partial}
%
%
%
\twocolumn[\hsize\textwidth\columnwidth\hsize\csname
@twocolumnfalse\endcsname

\title{
Properties of general relativistic, irrotational binary neutron stars 
in close quasiequilibrium orbits : Polytropic equations of state 
}

\author{K\=oji Ury\=u$^1$, Masaru Shibata$^{2,3}$ and Yoshiharu Eriguchi$^4$}
\address{$^1$SISSA, Via Beirut 2/4, Trieste 34013, Italy \\
$^2$Department of Physics, 
University of Illinois at Urbana-Champaign, Urbana, IL 61801, USA \\
$^3$Department of Earth and Space Science,~Graduate School of
 Science,~Osaka University,\\
Toyonaka, Osaka 560-0043, Japan \\
$^4$Department of Earth Science and Astronomy,
Graduate School of Arts and Sciences,
University of Tokyo, \\ Komaba, Meguro, Tokyo 153-8902, Japan}
\maketitle
\begin{abstract}
We investigate close binary neutron stars 
in quasiequilibrium states in a general relativistic framework.  
The configurations are 
numerically computed assuming (1) existence of a helicoidal 
Killing vector, (2) conformal flatness for spatial components of 
the metric, (3) irrotational velocity field for the neutron stars 
and (4) masses of neutron stars to be identical. 
We adopt the polytropic equation of state 
and the computation is performed for 
a wide range of the polytropic index $n~(=0.5, 0.66667, 0.8, 1, 1.25)$, 
and compactness of neutron stars $\compa~(=0.03 - 0.3$), where 
$M$ and $R$ denote the mass and radius of neutron stars in 
isolation. Because of the assumption of the irrotational velocity field, 
a sequence of fixed rest mass can be identified as 
an evolutionary track as a result of radiation reaction of 
gravitational waves.  Such solution sequences are computed from 
distant detached to innermost orbits where a cusp (inner Lagrange point) 
appears at the inner edges of the stellar surface. 
The stability of orbital motions and the gravitational wave frequency 
at the innermost orbits are investigated.  It is found that the 
innermost stable circular orbits (ISCO) appear for the case 
of stiff equation of state with $n \alt 2/3$. 
We carefully analyze the ISCO for $n=0.5$ 
and show that the ISCO are mainly determined by a hydrodynamic 
instability for $\compa \lo 0.2$.  
We also investigate the total angular momentum and the specific angular 
momentum distribution of the binary configuration at the innermost orbits, 
where the final merger process starts.  From these quantities, we 
expect the final outcomes of the binary neutron star coalescence.  
\end{abstract}
\pacs{PACS number(s): 04.25.Dm, 04.30.Db, 04.40.Dg, 97.60.Jd}
\vskip2pc]   
%
\baselineskip 5.mm

\section{Introduction}\label{intro}

The last $10$ minutes of binary neutron stars 
is one of the most important targets of laser interferometric 
gravitational wave detectors such as LIGO \cite{cutler,gwref}, 
as well as 
a possible candidate for various astrophysical phenomena such as 
$\gamma$-ray bursts \cite{piran} and optical transient events \cite{LP}.  
The close binary neutron stars are in quasi circular orbits and 
evolve as a result of gravitational wave emission 
towards merger. The evolutionary track of binary neutron stars is 
roughly divided into three phases; 
(i) inspiraling phase in which the characteristic 
orbital separation $a$ is much larger 
than the radius of neutron stars $R$ and the binary evolves 
adiabatically, 
(ii) intermediate phase in which the binary still evolves in 
the adiabatic manner but $a/R$ is so small ($\sim 2-5$) that the 
effects of tidal deformation of the neutron stars become important, 
and (iii) the final coalescing phase in which the two neutron stars 
merge on the dynamical timescale.  
The phase (i) has been treated by an analytic method using 
a post Newtonian approximation and equations of 
motion for point particles, resulting in 
recent successful developments \cite{blanchet}. 
On the other hand, phases (ii) and (iii) have to be treated numerically 
taking into account effects of the finite size, 
tidal deformation and hydrodynamic interaction of the two 
neutron stars. In this paper, we focus on phase (ii). 

The signal of gravitational waves in phases (ii) and (iii) 
is expected to give a wide variety of information about 
neutron stars and strong gravitational fields. 
In particular, the signal around 
the innermost stable circular orbit (ISCO) at which the orbit of 
binaries becomes dynamically unstable to starting to merge can bring  
important information about the structure of the neutron stars. 
At the ISCO, the signal is likely to 
change abruptly from a quasi periodic wave form to a 
burst wave form. In the frequency domain, 
such a change will be recognized as a clear cliff \cite{centrella}. 
The characteristic frequency at the edge of the cliff may be used for 
extracting information about the nature of 
equations of state of high density neutron star 
matter, because the location of the ISCO may depend sensitively on 
the equations of state (or compactness) of the neutron stars 
\cite{cutler}. 
The expected frequency for the ISCO, $\fgw$, can be estimated as 
\beq \label{freq}
\fgw = \frac{\Omega}{\pi} \simeq 
860\ {\rm Hz} \left(\frac{M_0\Omega}{0.02}\right) 
\left(\frac{M_0}{1.5M_\odot}\right)^{-1},  
\eeq
where $\Omega$, $M_0$ and $M_{\odot}$ are the angular velocity, 
the rest mass of each star and the solar mass.  
Thus, for a typical value of $M_0\sim 1.5M_{\odot}$, 
$\fgw$ may be smaller than 1kHz, which will be in the sensitive 
frequency band of laser interferometers \cite{footligo}, for 
a relatively small value of $M_0 \Omega \sim 0.02$ 
(see, {\it e.g.}, Tables II--III). This implies that 
the characteristic frequency may 
be detected by laser interferometers in the near future 
to constrain the equation of state of neutron stars. 
One of the main purposes of this paper is to investigate $\fgw$.

Other motivations for this paper are to prepare initial conditions 
for numerical simulations of merging of binary neutron stars 
in full general relativity as well as to 
predict possible outcomes in simulations starting 
from such initial data sets. 
Recently, several groups have been developing numerical codes 
for such simulation \cite{on97,ss99,sh99,su00,lt99}, and 
it is now becoming feasible to perform the simulation stably and 
fairly accurately (see, {\it e.g.}, \cite{su00}). 
As indicated in \cite{su00}, outcomes of the merger depend sensitively 
on initial velocity field and compactness 
of the neutron stars, so that 
we need to prepare the initial conditions as realistically as possible 
in order to obtain reliable results. 
It should also be emphasized that 
the simulation is not still an easy task 
and that it is very helpful to retain a reliable method 
to cross-check the results. 
Careful analysis of the initial data sets 
is one of the suitable methods for such cross-checking.

In this paper, we present numerical 
solutions for irrotational binary neutron stars of equal mass 
in quasiequilibrium states. 
We construct such states assuming the existence of 
a helicoidal Killing vector (see Eq.~(\ref{Killing})) because 
the emission timescale for gravitational waves is longer than 
the orbital period even just before merging and hence 
the system is almost in a stationary state 
in the comoving frame. 
We also assume an irrotational velocity field for the neutron stars 
because it is recognized as a realistic one for binary 
neutron stars as long as the spin period of the neutron stars 
is not as short as about 1 millisecond \cite{kbc92}. Under these 
assumptions, the relativistic Euler equation can be integrated 
to give a Bernoulli-type 
equation \cite{bas978,sh98,te98}, resulting in a great simplification 
for handling the matter equations.  

We adopt a polytropic equation of state in the form 
\beq \label{eos}
P=\kappa \rho^{1+1/n}
\eeq
where $P$, $\rho$, $n$ and $\kappa$ 
are the pressure, rest mass density, 
polytropic index, and polytropic constant, respectively.  
The adiabatic assumption ({\it i.e.}, $\kappa$ is a constant) 
is appropriate because the timescale for heating and cooling inside 
neutron stars is much longer than the inspiraling timescale of 
binaries due to gravitational radiation. 
Since realistic equations of state are not clear yet \cite{EOS}, 
we choose a wide range for $n$ between $n=0.5$ and 1.25 so that 
various moderately stiff equations of state 
for neutron star matter can be approximated. 
We also survey a wide range of the compactness parameter from 
$\compa = 0.03$ to $0.3$ 
where $\compa$ denotes the compactness of neutron stars in isolation. 
With these parameter sets, we will be able to explore 
the effects of various realistic equations of state appropriately.

This paper is organized as follows : In Sec.~\ref{sec2}, 
we briefly review our computational method and definition 
of physical quantities.  In Sec.~\ref{sec3}, 
we show numerical results. We analyze the quasiequilibrium 
states from various points of view. 
First, we construct a large number of quasiequilibrium sequences 
of constant rest mass for a wide range of $n$ and $\compa$. 
Then, we investigate the stability of orbital motions for these sequences. 
The stability is analyzed by searching the simultaneous minima of 
energy and angular momentum as a function of the 
orbital separation (or angular velocity) along each sequence 
\cite{lrs93,lrs97}. 
The mechanism of the instability and 
the gravitational wave frequency at the innermost orbits are determined. 
We also show that 
the maximum density decreases with decreasing orbital separation. 
Finally, we analyze the non-dimensional angular momentum parameter 
$q \equiv J_{\rm tot}/M_{\rm ADM}^2$, 
where $J_{\rm tot}$ and $M_{\rm ADM}$ 
are the total angular momentum and ADM mass of the system, and 
the distribution of the specific angular momentum 
of mass elements for binaries at the 
innermost orbits to predict possible outcomes after the merger. 
In Sec.~\ref{sec4}, we summarize the results presented 
in Sec.~\ref{sec3}.  In the Appendix, we demonstrate for completeness 
that neutron stars in the irrotational binary systems computed in 
this paper are stable against radial gravitational collapse.  
Throughout this paper, we take units in which $G=1=c$ where 
$G$ and $c$ are the gravitational constant and speed of light. 
Numerical computations are performed using spherical polar 
coordinates $(r, \theta, \varphi)$. 
Greek and Latin indices run over $t$(or 0), $r, \theta, \varphi$ and 
$r, \theta, \varphi$, respectively. 

\section{The method for constructing a quasiequilibrium sequence}
\label{sec2}

\subsection{Outline of formulation}

Orbits of binary neutron stars shrink as a result of radiation reaction 
to gravitational waves. The ratio of the radiation reaction 
timescale to the orbital period is evaluated to be \cite{st83}
\beq
\sim {5 \over 128\pi}\biggl({a \over M_{\rm ADM}} \biggr)^{5/2} \simeq
1.1 \biggl( {a \over 6 M_{\rm ADM}} \biggr)^{5/2}. \label{gwave}
\eeq
In deriving Eq. (\ref{gwave}), we 
adopted the quadrupole formula and the energy equation in  
Newtonian theory. 
Eq. (\ref{gwave}) implies that even just before the 
merger at $a \agt 6\madm$, the reaction timescale is still 
longer than the orbital period and that the binary is approximately 
in a quasi stationary state in the comoving frame 
with orbital angular velocity $\Omega$. 
Thus, neglecting the small effect of gravitational radiation 
reaction, we assume the existence of a helicoidal Killing vector in the form 
\beq
\ell^{\mu} = \biggl(\frac{\pa}{\pa t}\biggr)^{\mu} 
+ \Omega \biggl( \frac{\pa}{\pa \varphi} \biggr)^{\mu}. 
\label{Killing}
\eeq

Next, we assume that the velocity field of the neutron stars 
is irrotational. Viscosity inside neutron stars is considered 
to be too weak to synchronize the spin with the orbital rotation 
on the emission timescale for gravitational waves by the binary 
\cite{kbc92}.  In the final 
stage of inspiraling, the orbital period is about 2 milliseconds. 
Thus, if the spin period of the neutron stars 
is much longer than 2 milliseconds, 
the effect of the spin just before the merger is negligible so that the 
irrotational velocity field can be approximately achieved. 
In this assumption, furthermore, 
the relativistic Euler equation can be integrated 
to give a Bernoulli-type equation \cite{bas978,sh98,te98}. 
Consequently, procedure for obtaining a solution of the hydrodynamic 
equations is considerably simplified.

Finally, we assume that the spatial component of the metric 
$\gamma_{ij}$ is conformally flat and write the line element in the form 
\cite{wm956,bgm989,mmr99,ue99}
\begin{equation} \label{confo}
ds^2=-(\alpha^2 - \beta_i \beta^i)dt^2 + 2\beta_idx^i dt + 
\Psi^4 f_{ij} dx^i dx^j, 
\end{equation}
where $\alpha$, $\beta^i$, $\Psi$ and 
$f_{ij}$ are the lapse function, shift vector, 
conformal factor, and flat spatial metric, respectively. 
The elliptic-type equations for $\alpha$, $\beta^i$ and $\Psi$ 
are derived from the Hamiltonian constraint, momentum constraint 
and slicing condition in which $K_{ij}\gamma^{ij}=0$ with 
the assumption $\gamma_{ij}=\Psi^4 f_{ij}$ 
\cite{wm956,bgm989,mmr99,ue99} where $K_{ij}$ denotes the 
extrinsic curvature. 
The three metric is assumed to be conformally flat for simplicity. 
Although the solutions obtained in this framework are valid in 
general relativity in the sense that they 
satisfy the Hamiltonian and momentum constraints, 
they are approximate 
as quasiequilibrium states because of the 
simplified formulation. We suspect that the solutions 
shown below slightly and 
systematically deviate from the correct ones \cite{foot3}. 
It should always be kept in mind that 
we need to develop a better theoretical framework 
to suppress the slight deviation (see \cite{uue99} for one proposal).

Solutions of quasiequilibrium states of 
irrotational binary neutron stars are computed using a numerical code 
developed by Ury\=u and Eriguchi \cite{ue99}. 
The numerical method is described in \cite{ue98,ue99} 
to which the reader is referred for details.  
The code has been tested by comparing 
with numerical results of 
other groups ({\it e.g.}, \cite{bgm989}), as well as by carrying out 
several convergence tests \cite{commentcode}.

\subsection{Definition of variables}

First, we define the total rest mass $M_0$, 
ADM mass $M_{\rm ADM}$ and angular momentum $J_{\rm tot}$ as 
\beqn
M_{0,\rm tot} && = \int \rho \alpha u^0 \Psi^6 dV,\\
M_{\rm ADM} &&= \int \biggl[ \rho h (\alpha u^0)^2 -P 
+ {1 \over 16\pi}K_{ij}K^{ij} \biggr] \Psi^5 dV,\\
J_{\rm tot}&&=\int \rho h \alpha u^0 u_{\varphi} \Psi^6 dV,
\eeqn
where $u^{\mu}$ denotes the four velocity and $h=1 + (n+1)P/\rho$. 
The integral is carried out over the whole three space. 

We define 
the coordinate length of the semi-major axis $R_0$ 
and half of orbital separation $d$ as 
\beqn 
&& R_0 = (R_\ouou - R_\inin)/2 , \label{radi} \\ 
&& d  = (R_\ouou + R_\inin)/2 , \label{sepa}
\eeqn
where $R_\inin$ and $R_\ouou$ denote distances from 
the center of mass of the system to the inner and 
outer edges of the star respectively along the major axis.  
{}From these variables, we also define $\whd = d/R_0$ (note 
that $\whd=1$ when the surfaces come into contact and 
$\whd \rightarrow \infty$ for $d \rightarrow \infty$). 
In the following, we adopt $\whd$ to specify a model 
along a quasiequilibrium sequence. 

We also define half of the separation $d_G$ in another way as 
\begin{eqnarray} \label{d_G}
d_{G} =  \frac{1}{M_{0,\rm tot}}
\int | x | \rho \alpha u^0 \Psi^6 d V, 
\end{eqnarray}
where $x$ denotes a coordinate along the major axis.
Hereafter, we often refer to $d_G$ as the 
half of the orbital separation.

In this paper, we compute quasiequilibrium sequences fixing 
$\kappa$ to be constant. This treatment is 
appropriate for the late inspiraling stage of binary neutron stars  
because the timescale of heating and cooling is much longer than 
the inspiraling timescale of the binaries due to gravitational 
radiation reaction.  
Then, the physical units enter the problem only through the 
constant $\kappa$, which can be chosen arbitrarily or otherwise 
completely scaled out of the problem. Thus, 
we use non-dimensional quantities normalized by 
an appropriate power of $\kappa$ 
\beqn \label{paraR}
\bar{R}_0 &=& \kappa^{-n/2} R_0 ,  \\
\bar d_G &=& \kappa^{-n/2} d_G ,  \\
\bar M_{0,\rm tot} &=& \kappa^{-n/2} M_{0,\rm tot},  \\
\bar M_{\rm ADM} &=& \kappa^{-n/2} M_{\rm ADM}, \\
\bar J_{\rm tot} &=& \kappa^{-n} J_{\rm tot}, \\
\bar \rho &=& \kappa^{n} \rho. 
\eeqn
In the following, we often refer to a non-dimensional 
angular momentum parameter 
$q \equiv J_{\rm tot}/M_{\rm ADM}^2$. For convenience, 
half of the total rest mass, ADM 
mass and angular momentum are also defined as $M_0 = M_{0,\rm tot}/2$, 
$M = M_{\rm ADM}/2$ and $J = J_{\rm tot}/2$ 
with their normalized values $\bar M_0=\kappa^{-n/2}M_0$, 
$\bar M  =\kappa^{-n/2}M$, and $\bar J=\kappa^{-n}J$. 
Strictly speaking, $M$ is not 
the mass of one star because the binding energy between two bodies 
and other interaction effects are included in it. However, the 
fraction is expected to be fairly small, so that 
we often refer to $M$ as the mass of each star.

\section{Results}
\label{sec3}

\subsection{Stability of quasiequilibrium sequences 
of binary neutron stars}
\label{sec3a}

Since the vorticity is almost exactly conserved during the evolution 
of binary neutron stars, 
a quasiequilibrium sequence of irrotational binary neutron stars of 
constant rest mass and with decreasing the orbital separation 
(or increasing $\Omega$) 
can be identified as an evolutionary sequence of 
binary neutron stars as a result of gravitational wave emission.  
We compute the sequences for a 
wide range of parameter space as $0.5 \leq n \leq 1.25$ 
and typically as $0.1 \leq \compa \leq 0.19$. 
(Hereafter, we use $\compa$ instead of the rest mass 
to specify a certain sequence.) 
In some cases, we extend the analysis to $\compa < 0.1$ and 
$\compa \geq 0.2$. 
In this paper, we restrict our attention only 
to binary neutron stars of equal mass.

Each sequence is computed gradually decreasing $\whd$ from 3 to 1.  
It is found that the sequences are always terminated at an innermost 
orbit with $\whd > 1$ and $\bar d_G=\bdR$ 
at which neutron stars have cusps at the inner edges of the 
stellar surfaces. 
This property is found irrespective of $n$ and $\compa (\alt 0.2)$. 
The cusps correspond to the inner Lagrange (L1) points.  
Therefore, neutron stars at $\bar d_G=\bdR$ 
are subject to mass transfer from 
the cusps and are likely to form a dumbbell-like structure 
({\it i.e.}, the system will have 
a bridge between the two stars and will not 
be a ``binary'' any longer for $\bar d_G < \bdR$). 
Even at this orbit of cusps, the 
timescale of gravitational radiation reaction is longer 
than the orbital period, except for extremely compact 
binaries, so that such a dumbbell-like object could be still in a 
quasiequilibrium state and may be computed with the same strategy 
as used here. However, 
we stop the computation at $\bar d_G=\bdR$ since 
the present numerical code is suitable only for binary 
configurations.  
We show several physical quantities at this orbit in Table I.

To determine the dynamical stability of the orbit of binaries, 
we search for minima of $\bJ$ and $\bM$ as 
functions of the orbital separation (or the angular velocity) 
along each sequence.  
Lai, Rasio, and Shapiro \cite{lrs93} have shown the 
existence of minima of $\bJ$ and $\bM$ 
for an ellipsoidal model of a Newtonian irrotational binary, 
which indicates the onset of dynamical instability.  
(See Lombardi, Rasio and Shapiro \cite{lrs97} for an extension 
including a part with post Newtonian corrections.)  These minima of 
$\bJ$ and $\bM$ should appear simultaneously at the same separation 
because the following relativistic identity probably holds 
for irrotational sequences with constant rest mass; 
\beq
d\madm = \Omega dJ_{\rm tot}.  
\eeq
In our numerical computation, this identity is satisfied typically to 
$\lo 10 \%$ except near to turning points where 
$\bM$ and $\bJ$ are almost constant and consequently the 
difference between the neighborhoods is not accurately computed. 
Also, we checked that the minima of $\bJ$ and $\bM$ appear 
simultaneously at the same separation $\bdg=\bddy$ whenever 
they appear on a solution sequence \cite{commentsepa}.  
In view of this, we define 
the location of the simultaneous minima of $\bJ$ and $\bM$ 
as the ISCO.

\begin{figure}
\epsfxsize=8truecm
\begin{center}
\epsffile{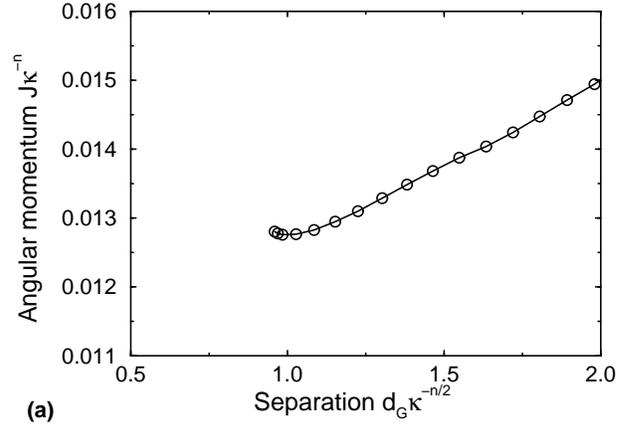}
\end{center}
\epsfxsize=8truecm
\begin{center}
\epsffile{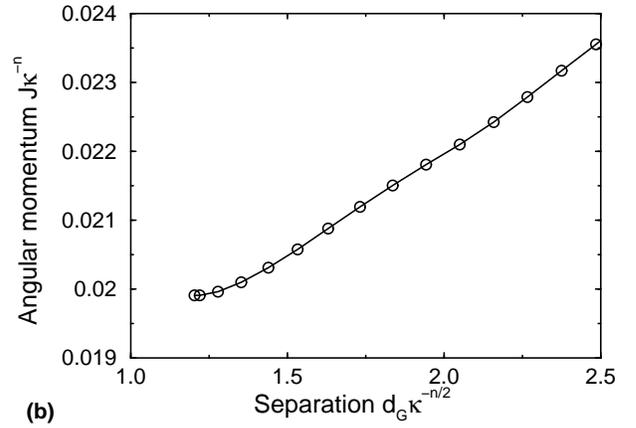}
\end{center}
\epsfxsize=8truecm
\begin{center}
\epsffile{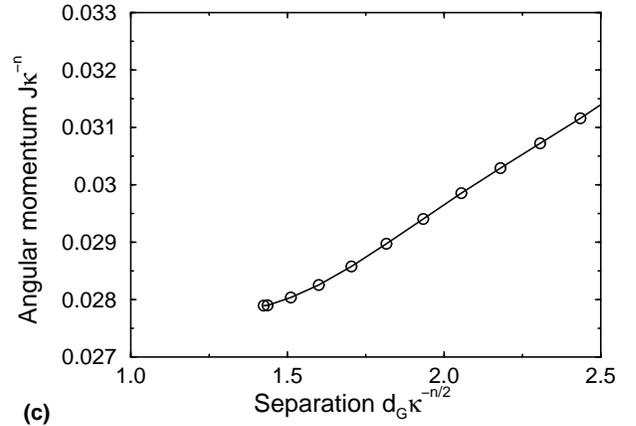}
\end{center}
\caption{
$\bar J$ as a function of $\bar d_G$ for (a) $n=0.5$, 
(b) $n=0.66667$ and (c) $n=0.8$ and for $\compa = 0.19$.  
The points of smallest $\bdg$ along the curves correspond 
to $\bar d_G=\bdR$ where neutron stars have cusps at the  
inner edges of the stellar surfaces. 
\label{fig1}}
\end{figure}

In Fig.~\ref{fig1}, we show $\bar J$ as a function of $\bar d_G$ 
for $n=0.5$, $0.66667$ and $0.8$, and 
for $\compa = 0.19$. (Note that we have found essentially the same 
behavior for $0.1 \leq \compa \leq 0.19$.) 
For $n=0.5$, $\bar J$ and $\bar M$ 
have the minima at $\bar d_G=\bddy > \bdR$ and so 
the orbits of the binary neutron stars will be dynamically 
unstable to merger before mass transfer sets in. 
For $n=0.66667$, the minima are located near $\bdR$ 
and so dynamical instability and mass transfer will set in almost 
together.  
On the other hand, for $n \geq 0.8$, 
minima are not found for $\bar d_G \geq \bdR$ 
and so the mass transfer will set in before the orbit becomes 
dynamically unstable.

Several quantities at $\bar d_G=\bddy$ for $n=0.5$ and 0.66667 are 
summarized in Tables II and III, respectively. 
As shown in Table I, $\whd$ at $\bdg = \bdR$ depends very weakly on 
$\compa$.  In contrast, 
$\whd$ at $\bdg = \bddy$ becomes larger with increasing $\compa$, 
as shown in Tables II and III.  
For $n=0.5$, an ISCO always exists for any $\compa \leq 0.3$.  
For $n=0.66667$, no ISCO exists for the sequence of 
$\compa < 0.17$, but there is one for $\compa \go 0.17$. 
(Note that the curves of $\bJ$ and $\bM$ as functions of
$\bdg$ suggest that 
$\bdR$ is close to $\bdg$ even for $\compa < 0.17$ for $n=0.66667$.)  
The results for the smaller compactness $\compa < 0.17$ with 
various $n$ are consistent 
with the result in the Newtonian limit $\compa \rightarrow 0$ 
\cite{ue98}, implying that the above property of the dynamical 
instability holds 
irrespective of compactness as long as $\compa < 0.17$.

\begin{figure}
\epsfxsize=8truecm
\begin{center}
\epsffile{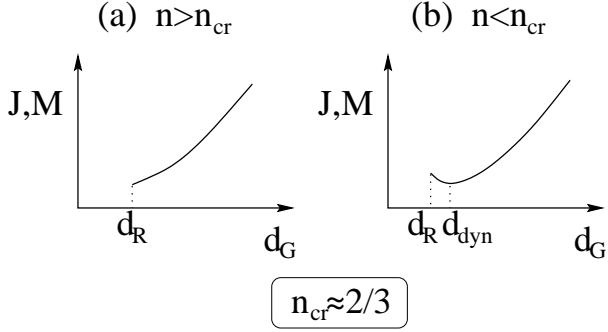}
\end{center}
\caption{Schematic figure of $J$ and $M$ as functions of $d_G$ 
for sequences of constant rest mass.  
$\dR$ corresponds to the point where neutron stars have cusps 
at the inner edges of the stellar surfaces. 
$\ddy$ corresponds to a turning point along the sequence. 
\label{fig2}}
\end{figure}

To summarize, we show a schematic figure in Fig.~\ref{fig2}.  
We have found two cases for evolution of close binary neutron stars 
with irrotational velocity field. 
In one case, the sequence is terminated when cusps appear before 
the binaries reach 
the minima of $\bar J$ and $\bM$ as shown in Fig.~\ref{fig2}(a). 
In this case, mass transfer will occur before the orbit 
becomes dynamically unstable. 
Binary neutron stars with $n \agt 2/3$ are in this category. 
In the other case,  the binaries reach the minima 
before the cusps appear as shown in Fig.~\ref{fig2}(b).  
In this case, the orbit becomes dynamically unstable 
before mass transfer sets in.  
Binary neutron stars with $n \alt 2/3$ are in this category. 
Based on this result, we hereafter refer to the orbit 
at $\bar d_G=\bddy$ for $n\alt 2/3$ and at $\bar d_G = \bdR$ for 
$n \geq 0.8$ as ``innermost orbit''. 

\subsection{Frequencies of gravitational waves at the final orbit}
\label{sec3b}

As discussed in Sec.~I, it is important to clarify 
the frequency of gravitational waves at the ISCO from the viewpoint of 
gravitational-wave-astronomy. 
For $n=0.5$ and 0.66667, we find that an ISCO ({\it i.e.}, 
a point at which dynamical instability sets in) does exist, and 
the frequency is computed from Tables II and III as 
\beqn
&&\fgw \nonumber \\
&&\simeq \left\{
\begin{array}{ll}\displaystyle 
0.8 \biggl( {1.5M_{\odot} \over M_0}\biggr) {\rm kHz}
& {\rm for}~\compa = 0.14, \\
\displaystyle 1.1-1.15\biggl( {1.5M_{\odot} \over M_0}\biggr) {\rm kHz}
& {\rm for}~\compa = 0.17, \\
\displaystyle 1.35-1.4\biggl( {1.5M_{\odot} \over M_0}\biggr) {\rm kHz}
& {\rm for}~\compa = 0.19. \\
\end{array}
\right. 
\label{fisco}
\eeqn

We note that the ISCO here for $\compa \alt 0.2$ is 
probably determined 
by the hydrodynamic instability \cite{lrs93}, but not by the general 
relativistic orbital instability.  The reason is that 
the general relativistic orbital instability should not depend 
on $\compa$ in the absence of hydrodynamic effects \cite{footisco}.  
Even in the presence of such effects, $\madm \Omega$ should depend 
only weakly on 
$\compa$ when the general relativistic orbital instability 
dominates to determine the ISCO.  
However, for $\compa \alt 0.2$, $M_{\rm ADM}\Omega$ at the ISCO 
does depend strongly and systematically on $\compa$ (cf, Tables II 
and III, and Fig.~\ref{figMO}). 
We will discuss this point in detail in Sec.~\ref{sec3c}.

For a realistic neutron star with $M_0 \simeq 1.5M_{\odot}$, the 
radius will be in the range between 10 and 15km. This implies that 
$\compa$ is between $0.14 \lo \compa \lo 0.2$ for a typical ADM mass 
$M_{\infty} \sim 1.4M_{\odot}$ where $M_{\infty}$ denotes the ADM 
mass in isolation. 
The present result suggests that if the neutron star radius is 
fairly large, {\it i.e.}, $\sim 15$km, the frequency of gravitational 
waves at 
the ISCO will be less than 1kHz which is in the 
sensitive frequency band of 
the laser interferometers. However, 
if the radius of neutron stars is $\sim 10$ km, 
the frequency at the ISCO is larger than 1kHz and 
it will be difficult to detect gravitational waves emitted from 
binary neutron stars with $M_{\infty} \sim 1.4 M_{\odot}$ orbiting 
near the ISCO by means of the interferometers \cite{footligo}.

For $n=0.8$ and 1, the frequency of gravitational waves 
at $\bar d_G = \bar d_R$ is given 
approximately by Eq.~(\ref{fisco}), implying that 
$\fgw$ would be larger than the frequency shown in Eq.~(\ref{fisco}) for 
these cases. Hence, even for $R \sim 15$km and $M_0 \sim 1.5M_{\odot}$, 
the frequency at ISCOs may be larger than 1kHz.

It should be pointed out that for $n \geq 0.8$, 
mass transfer is likely to set in before the binaries 
reach the ISCO as discussed above. If the timescale of mass transfer 
is longer than the timescale for evolution due to gravitational 
wave emission, the signal of the gravitational waves may not change 
abruptly from a quasi periodic type to a different one at this orbit.  
However, if the mass transfer proceeds quickly, 
a characteristic signal may be produced. 
As mentioned above, 
the frequency of gravitational waves at the onset of mass transfer 
is approximately given by Eq.~(\ref{fisco}) for $n=0.8$ and 1. 
Thus, for the case of relatively soft equations of state, 
the characteristic signal at mass transfer may be detected 
for binary neutron stars of $\compa \sim 0.14$ 
and $M_{\infty} \sim 1.4 M_{\odot}$.

\subsection{Origin of the ISCO} 
\label{sec3c}

\begin{figure}
\epsfxsize=8truecm
\begin{center}
\epsffile{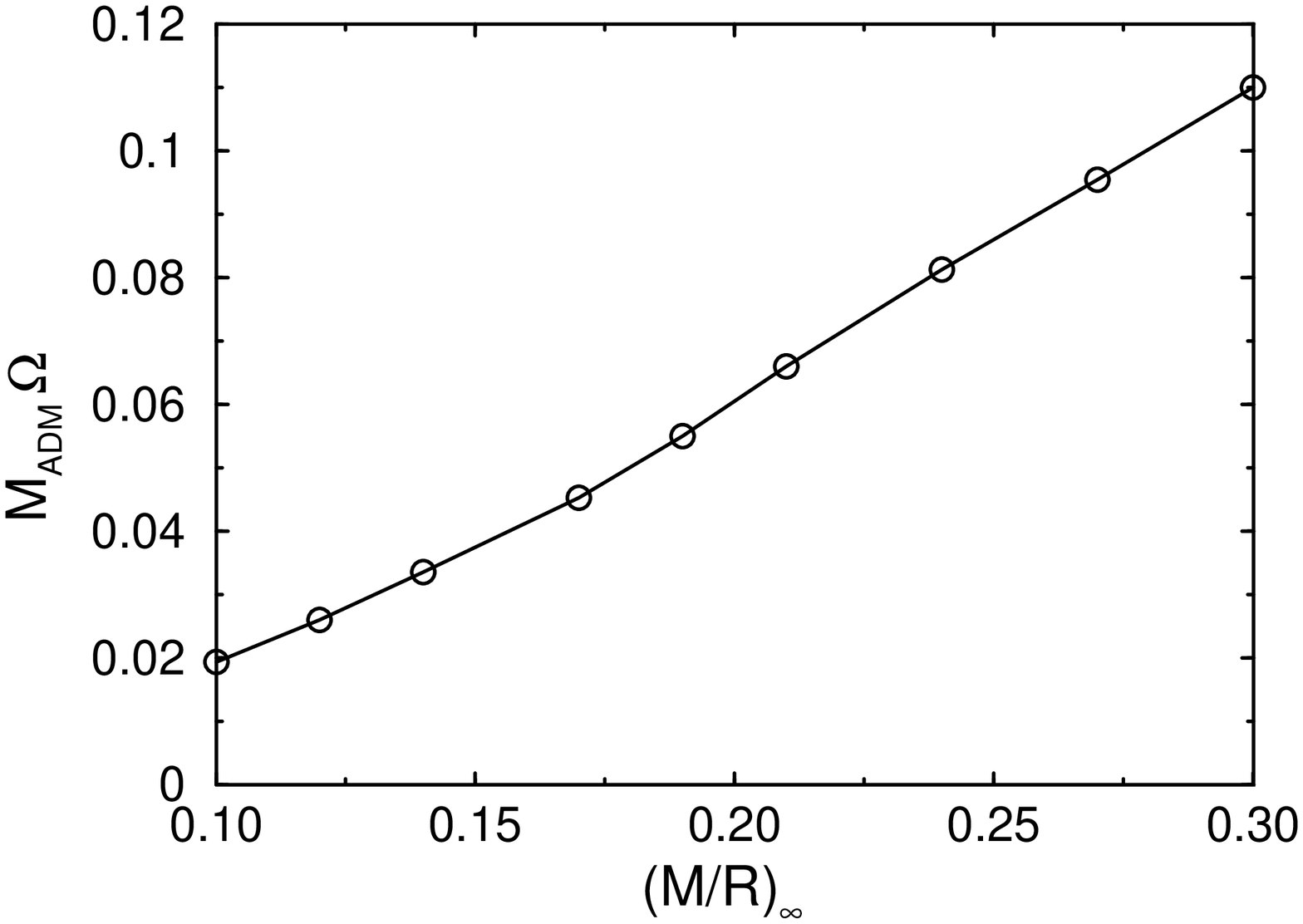}
\end{center}
\caption{
Plot of $\compa$ -- $\Omega \madm$
relations at the ISCO $(\dg \approx \ddy)$
for binary neutron stars with $n=0.5$ and $n=0.66667$.  
\label{figMO}}
\end{figure}

In Fig.~\ref{figMO}, we plot $\madm \Omega$ at ISCOs for 
$n=0.5$ as a function of $\compa$ using data tabulated 
in Table II.  
As shown in Fig.~\ref{figMO} and as argued in Sec.~\ref{sec3b}, 
$\madm \Omega$ at the ISCOs increases with increasing $\compa$. 
This strongly indicates that the ISCO is determined by the 
hydrodynamic instability \cite{lrs93}. 
On the other hand, 
in the limit $\compa \rightarrow 1/2$, 
hydrodynamic effects should become less important so that the 
general relativistic orbital instability can determine the ISCO. 
In our present formulation, 
not only hydrodynamic but also 
general relativistic effects (in the framework of the 
conformal flatness approximation) are taken into account, 
which enables us to investigate the origin of the ISCO.   
In this subsection, we carry out a detailed analysis 
using the results of $n=0.5$.

As a first step, we derive fitting formulae for 
$\compa$ $<$ $0.2$ to clarify the behavior of $M_0\Omega$ and 
$M\Omega$ at ISCOs as functions of $\compa$. 
In the derivation, the following two points are 
taken into account: (1) In the Newtonian limit $\compa \rightarrow 0$, 
$M_0 \Omega \compa^{-3/2}$ and $M \Omega \compa^{-3/2}$ converge 
to an identical constant 
because they are non-dimensional and should be 
identical in Newtonian gravity. 
(2) In the Newtonian limit, the ISCO is determined by the 
hydrodynamic effect \cite{lrs93}, and this is the case 
for the finite $\compa$ as long as $\compa$ is small. 
Consequently, all of the general relativistic corrections should 
appear in the power series of $\compa$ 
from the viewpoint of the post Newtonian approximation. 
{}From (1) and (2), we fix the functions of $M_0\Omega$ and 
$M \Omega $ in the form 
\beqn \label{fiteq}
&&M_0\Omega \compa ^{-3/2}=a + b\compa + c\compa^2,\\
&&M \Omega \compa ^{-3/2}=a' + b'\compa + c'\compa^2, 
\eeqn
(namely, we expand them up to second post Newtonian order) 
and determine the coefficients $(a, b, c)$ and $(a', b', c')$ 
by the least square fitting. 
We here note that although we know the value for $a=a'$ 
from the Newtonian computation \cite{ue98}, 
we do not use the data for the fitting.  Instead, we compute the 
value at $\compa = 0$ from the fitting formula and compare with 
the Newtonian results for a cross-checking. 

In the least squares fitting, we use 
the data sets for $n=0.5$ of 
$0.03 \leq \compa \leq 0.17$ (or $0.19$) shown in Tables II.  
The resulting coefficients for the fitting formulae 
are tabulated in Table IV.  
We may roughly conclude, 
\beqn
a&=&0.286 \pm 0.001, \quad b=0.40\pm 0.03, \nonumber \\
a'&=&0.286 \pm 0.001, \quad b'=0.22\pm 0.02. 
\eeqn
For $c$ and $c'$, on the other hand, 
a large uncertainty of $O(0.1)$ exists. 
In a Newtonian analysis,  
$M_0\Omega \compa^{-1.5}$ $=M\Omega \compa^{-1.5}$ $ \simeq 0.284$ 
for $n=0.5$ \cite{ue98}, which 
agrees well with $a$ and $a'$ in the above fitting formulae.
This shows the validity of the fitting method adopted here. 

\begin{figure}
\epsfxsize=8truecm
\begin{center}
\epsffile{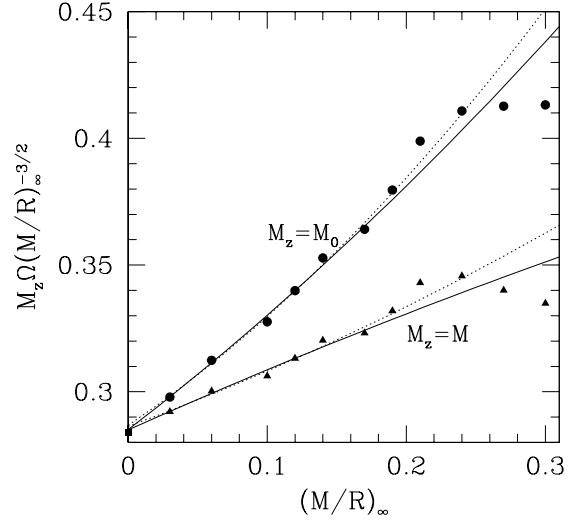}
\end{center}
\caption{
$M_0 \Omega$ (filled circles) and $M\Omega$ (filled triangles) 
at the ISCO $(\dg \simeq \ddy)$ 
as a function of $\compa$ for $n=0.5$.  
The solid and dotted lines denote the fitting formulae derived 
using the data sets with $0.03 \leq \compa \leq 0.17$ (solid lines) 
and $0.03 \leq \compa \leq 0.19$ (dotted lines), respectively. 
The squares at $\compa = 0.0$ denote the results from the Newtonian 
computation. 
\label{fig4}}
\end{figure}

In Fig.~\ref{fig4}, 
$M_0 \Omega \compa^{-1.5}$ and 
$\madm \Omega \compa^{-1.5}$ at the 
ISCO are plotted as a function of $\compa$ together with 
results from the fitting formulae derived above.
The filled circles and triangles denote the numerical results, and 
the solid and dotted lines denote results obtained from 
the fitting formulae. 
These figures indicate that the fitting formulae agree fairly well 
with numerical results 
for $\compa \alt 0.2$, implying that the post Newtonian expansion 
is reasonable in this regime. 
In particular, the fitting is very good around 
$\compa \alt 0.15$.  
In this regime, $M_0 \Omega$ and $\madm \Omega$ are well approximated 
by only the first post Newtonian correction, {\it i.e.}, 
the $\compa^2$ term is not very important.

The results of the fitting imply that the frequency of the ISCO depends 
strongly and systematically on $\compa$. Taking into account that 
the ISCO in the Newtonian limit is determined by the 
hydrodynamic instability, we may naturally infer that the ISCO 
for small $\compa \alt 0.2$ 
is still determined by the hydrodynamic instability 
with general relativistic corrections. 
As we mentioned in Sec.~\ref{sec3b}, the compactness of a realistic 
neutron star is likely to be in the range between 
$0.14 \alt \compa \alt 0.2$. Therefore, 
the ISCO of the realistic binary neutron stars 
is determined by the hydrodynamic instability but not by the 
general relativistic instability.

\begin{figure}
\epsfxsize=8truecm
\begin{center}
\epsffile{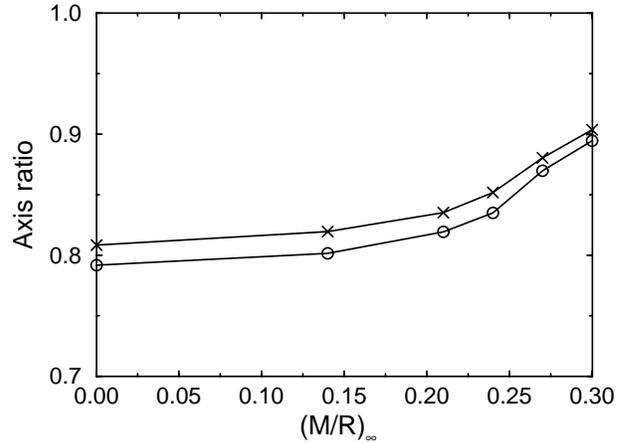}
\end{center}
\caption{
The ratios $a_2/a_1$ (circles) and $a_3/a_1$ (crosses) 
as functions of $\compa$ at ISCOs for $n=0.5$.  Values at 
$\compa = 0.0$ are those from the Newtonian computation. 
\label{figaxisrat}}
\end{figure}

For $0.2 < \compa \leq 0.3$, on the other hand, the fitting formulae 
do not agree very well with the numerical results, implying 
that the angular velocity at the ISCOs 
seems to be affected by strong nonlinear 
effects of general relativity.  
In this regime, the hydrodynamic effects seem to be less important.  
In Fig.~\ref{figaxisrat}, we show the ratio of semi-diameters for 
neutron stars at the ISCOs as a function $\compa$ for $n=0.5$. 
The semi-diameters $(a_1, a_2, a_3)$ are computed from 
\beq
a_n = \int \Psi^2 d \ell  \quad (n=1,2,3),
\eeq
where the line integral is taken along straight lines in 
Cartesian coordinates, and 
the three directions ($n=1-3$) are chosen to be orthogonal each other. 
$a_1$ is the longest diameter along a line connecting the centers of 
the two neutron stars.  $a_2$ and $a_3$ are those in the equatorial 
and the meridional plane, respectively, and they intersect each other 
at the coordinate center of $a_1$.  
The circles and crosses 
in Fig.~\ref{figaxisrat} denote $a_2/a_1$ and $a_3/a_1$, respectively. 
These ratios appear to converge towards 
unity with increasing $\compa$. 
This indicates that the tidal effect is less important for determination 
of the ISCO for $\compa \sim 0.3$. 
However, even in this highly general relativistic regime, 
$\madm \Omega$ still does not converge to a constant as 
shown in Fig.~\ref{figMO}. 
Since $\madm \Omega$ depends strongly on $\compa$,  
the effect of the hydrodynamic instability still appears to dominate 
over the general relativistic orbital instability for 
determining the ISCO.

Cook \cite{co94} and recently Baumgarte \cite{B00} 
have investigated the ISCO for binary black holes using 
the conformal flatness approximation for the geometry as we do.  
According to their results, 
${\cal M}\Omega$ at the ISCO is 0.17--0.18. 
(${\cal M}$ is the sum of the gravitational masses 
of the two black holes, and is slightly different from $\madm$ 
although the difference is not important here.) 
As shown in Fig.~\ref{figMO}, $\madm \Omega \simeq 0.11$ at 
the ISCO for $\compa=0.3$ and it is still monotonically increasing.  
This tendency does not contradict their results.  
The ISCO in their work is purely determined 
by the general relativistic orbital instability. 
Therefore, we may expect that 
the ISCO of binary neutron stars 
is eventually determined by the general relativistic 
orbital instability for a sufficiently large $\compa > 0.3$ 
and $\madm \Omega$ converges to a constant $\sim 0.17-0.18$, 
although such a large compactness is 
unrealistic for neutron stars. 
(We have the value $\madm \Omega \simeq 0.18$ for 
$n=0.5$ case, if we naively extrapolate data points 
at the ISCO with $\compa = 0.24$ and $0.3$ linearly to 
$\compa = 4/9$.)

Finally, we compare the present results with ones computed 
using other analytic methods.  Kidder, Will and Wiseman \cite{kww92}, 
Damour, Iyer and Sathyaprakash \cite{dis98}, and Buonanno 
and Damour \cite{bd99} have determined the ISCO using relativistic 
equations of motion for two point particles 
in which some general relativistic corrections have been 
taken into account.  
They have derived the orbital angular velocity at the ISCO as 
${\cal M} \Omega \simeq 0.0605$, $0.08850$ and $0.07340$, 
respectively. 
Since equations of motion for point particles have been used 
in their methods, the ISCO has been determined 
by the general relativistic orbital instability and 
${\cal M} \Omega$ has been independent of $\compa$. 
This is in contrast with our results, in which $\madm \Omega$ 
depends on $\compa$ even around $\compa \sim 0.3$ and 
consequently the general relativistic orbital instability 
does not seem to dominate.  

The reason for this contradiction is clear for $\compa$ $\alt$ $0.2$. 
As we showed above, the ISCO for $\compa \alt 0.2$ is determined 
by the hydrodynamic instability. In this regime for the compactness, 
$\madm \Omega \alt 0.06$ which is smaller than their results, namely, 
that before the general relativistic effects between two stars 
become important, the orbits become unstable because of 
the hydrodynamic instability. 
Since realistic neutron stars have $0.14 \alt \compa \alt 0.2$, 
the approaches using equations of motion for point particles are 
not appropriate for investigating the ISCO of binary neutron stars.

For $0.2 < \compa \leq 0.3$ and $\madm \Omega \leq 0.11$, 
our present results suggest that the ISCO may be determined 
by the hydrodynamic 
instability. According to the point particle approaches, however, 
the general relativistic orbital instability should be important 
for $\madm \Omega >0.06-0.09$ implying that 
our results contradict theirs in this regime, too. 
The reason for this contradiction is not clear. 
The works of ourselves, Cook and Baumgarte 
are based on the conformal flatness 
approximation which may fail to include some important 
general relativistic effects \cite{foot3}. 
The neglected effects might play an important role in determining 
the ISCO for the regime $\compa > 0.2$. Thus, we should not draw 
any strong conclusions for such highly relativistic binaries 
from results obtained in this framework since this might be 
the reason for the contradiction. 
To pin down the uncertainty for the high compactness regime, 
we need to carry out computations 
including higher general relativistic corrections.

\subsection{Evolution of the maximum density}
\label{sec3d}

\begin{figure}
\epsfxsize=8truecm
\begin{center}
\epsffile{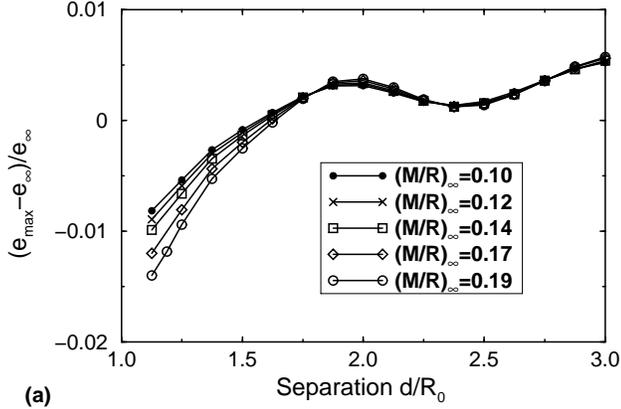}
\end{center}
\epsfxsize=8truecm
\begin{center}
\epsffile{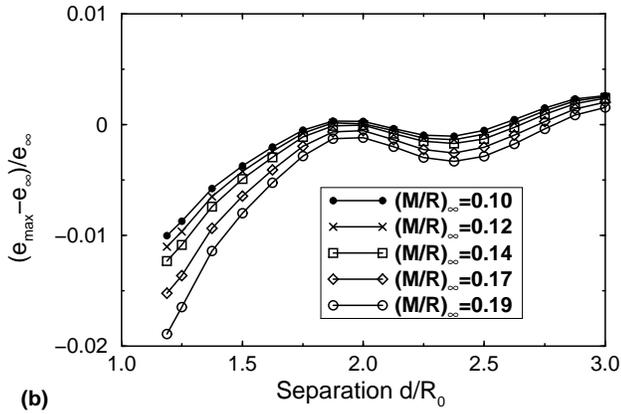}
\end{center}
\epsfxsize=8truecm
\begin{center}
\epsffile{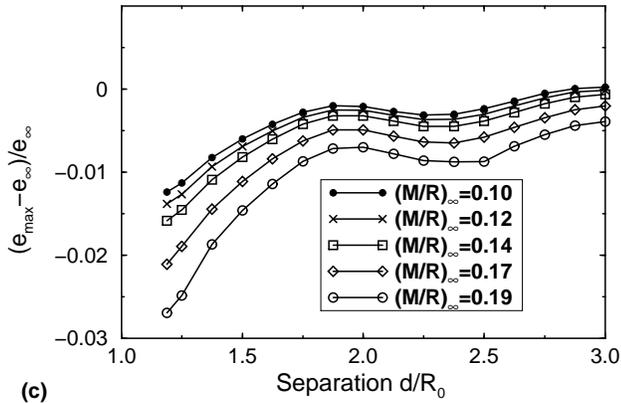}
\end{center}
\caption{
$\eee$ 
as a function of $\whd$ for (a) $n=0.5$, 
(b) $n=0.66667$ and (c) $n=0.8$.  
Cusps appear at the inner edges of the stellar surfaces 
at the point of smallest separation on each curve. 
\label{fig6}}
\end{figure}

In Figs.~\ref{fig6}(a)--(c), 
we show the relative change in the maximum energy 
density of a star 
$\Delta e\equiv (e_{\rm max} - e_\infty)/e_\infty$ 
as a function of 
$\whd$ for $n=0.5, 0.66667$ and $0.8$, respectively. 
Here, $e = \rho + n P$, and  $e_{\rm max}$ and $e_\infty$ 
denote respectively its maximum value for $\whd < \infty$ 
and its value for an isolated spherical star ($\whd = \infty$) 
as computed from the TOV equation.

The figures indicate that the maximum density slightly decreases 
with decrease of the orbital separation 
irrespective of $n$ for $\whd \lo 2 $. 
For larger separations of $\whd \go 2 $, 
each star becomes almost spherical so that $\Delta e$ 
can be less than the numerical uncertainty. 
The small global shifts $(< 1 \%)$ of each curve from zero for 
$2 \lo \whd \lo 3$ are thought to be due to numerical errors 
associated with our numerical scheme (a finite difference scheme).  
Small deviations around the average 
for $2 \lo \whd \lo 3 $ are due to a systematic error 
associated with the Legendre expansion for the 
gravitational field.  (See \cite{ue99} for details.)
Thus, our computation indicates no evidence that neutron stars in 
binary systems become dynamically unstable against gravitational 
collapse.  Instead, the maximum density slightly 
decreases with decreasing orbital separation 
irrespective of $n$. (See the Appendix for more detailed 
computations for selected values of $n$.)  
This result is consistent with recent analytic calculations in 
general relativity \cite{KIPeanna}.

\subsection{Prospects for the outcome after the merger}
\label{sec3e}

First, we investigate the quantity 
$q \equiv J_{\rm tot}/M_{\rm ADM}^2$ which 
is an important parameter for predicting the outcome 
after a merger. If the $q$ of the object resulting from 
the merger is larger than unity, it will not be able to collapse 
to form a black hole. 
Thus, if $q$ is larger than unity for a binary at its
innermost orbit, there is a possibility that 
no black hole will be formed promptly after the merger.  
On the other hand, if $q$ is less 
than unity before the merger, we could consider that the system is a 
candidate to become a black hole, because $q$ will not increase 
during the merger due to emission of gravitational waves \cite{footq}. 

\begin{figure}
\epsfxsize=8truecm
\begin{center}
\epsffile{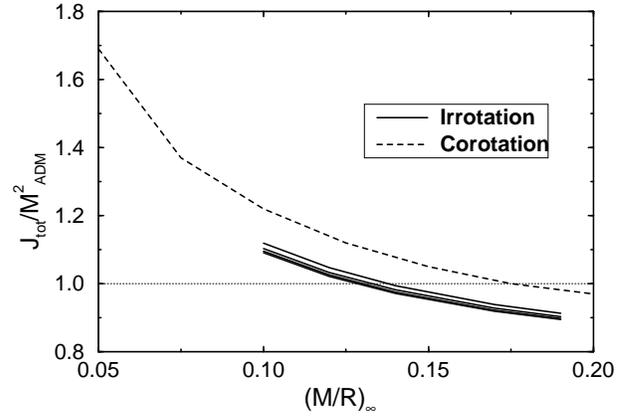}
\end{center}
\caption{
Relation between $\compa$ and $\jom$ 
at the innermost orbits of binary neutron stars. 
The solid lines denote the relation for irrotational binaries with 
$n=0.5$, 0.66667, 0.8 and 1 from top to bottom, and 
the dashed line is for corotating binaries with $n=1$. 
For the irrotational case, 
the models at $\bar d_G=\bar d_R$ are chosen as the innermost 
orbits, and for the 
corotating case, the models at the energy and angular momentum minima 
are chosen. 
\label{fig7}}
\end{figure}

In Fig.~\ref{fig7}, we show the relation between 
$\compa$ and $q$ at the innermost orbits 
of irrotational binary neutron stars for $n=0.5-1.25$ and 
we also show the relation for {\it corotating} binary neutron 
stars with $n=1$ for comparison.  
Here, the innermost orbits for irrotational binaries are 
the orbits at $\bar d_G=\bddy$ for $n = 0.5$ 
and at $\bar d_G=\bdR$ for $n \geq 2/3$ as we define in Sec.~\ref{sec3a}.  
(Note that $\bdR \simeq \bddy$ for $n = 2/3$.)  
For corotating binaries, we choose the orbit 
at the energy and angular momentum minima. 
It is found that 
$q$ is always less than unity irrespective of $n$ 
for irrotational binaries with $\compa \geq 0.13$.  
As mentioned in Sec.~\ref{sec3a}, $\compa$ will be larger 
than $\sim 0.14$ for a realistic neutron star. Therefore, 
$q$ for realistic binary neutron stars 
decreases below unity before merger, implying that the binary system 
satisfies a necessary condition to form a black hole. 

Here, two cautions are appropriate. (1) For very stiff equations of 
state, the total rest mass of the system can be smaller than the maximum 
allowed rest mass of a spherical star. Indeed, 
for $n=0.5$ and $\compa=0.14$, this is the case. 
Thus, for very stiff equations of state, a black hole will not be 
formed even if $q < 1$ before merger and $\compa$ is fairly 
large ($\sim 0.14$).  
(2) Even in the case 
when the system is supramassive in the sense that the rest mass 
of the system is larger than the maximum rest mass of 
a spherical star with the same equation of state, 
the condition $q < 1$ is not 
sufficient for black hole formation, because 
rapid rotation could support the self-gravity of the supramassive star 
\cite{BSS}.  To find the outcome of the merger, numerical 
simulations are obviously necessary.

In contrast with the above results, 
$q$ is larger than unity even for $\compa \sim 0.17$ at the 
energy and angular momentum minima for corotating binary neutron 
stars with $n=1$ \cite{ba97}. 
The reason for the difference between the two cases 
is that stars in the irrotational binaries 
have only negligible spin angular momentum while those 
in the corotating binaries have significant spin angular momentum.  
In the study of corotating binary neutron stars with 
$\compa \sim 0.17$, 
one could reach the conclusion that a black hole 
might not be formed after the merger, which is completely 
different from our present conclusion for 
irrotational binary neutron stars. 
This result brings us to another caution, that 
if we adopted corotating velocity fields, 
we could reach an incorrect conclusion 
for realistic binary neutron stars which will have (nearly) 
irrotational velocity fields.  

Next, we investigate the mass spectrum with respect to 
the specific angular momentum of fluid elements in order to 
estimate the mass of the disk around a black hole formed 
after merger. We consider this for binaries at innermost orbits 
which are chosen in the same manner as above. 
The mass spectrum with respect to the specific angular momentum 
is defined as 
\beq \label{rmfra}
\frac{M_0(j)}{M_0} = 
\frac{1}{M_0}
\int_{j'>j} \rho \alpha u^0 \Psi^6  d V, 
\eeq
where the integral is performed for fluid elements which have  
specific angular momentum larger than a given value $j$.  
The specific angular momentum $j$ is defined as 
\beq \label{speang}
j = h u_{\varphi}.
\eeq
Here, we note that $j$ of each fluid element is exactly conserved 
in axisymmetric systems for ideal fluid. 
For non-axisymmetric systems, it may not be conserved due to 
some dissipation processes such as gravitational radiation or 
other outward transportation process. 

In Figs.~\ref{fig8}, 
we show $\rmfrac$ as a function of $j/\madm$ 
for $n=0.5-1.25$, and $\compa=0.14$ and 0.17. 
Here, we choose binaries with $\compa \geq 0.14$ 
in order to be realistic.  
The vertical dotted lines in Figs.~\ref{fig8} denote $j/\madm$ of 
a test particle orbiting at the ISCO around a Kerr black hole for 
$q=0.99$, $ 0.95$, and $0.9$ \cite{st83}, 
showing the minimum allowed values for test particles orbiting 
a black hole. 
For $n = 1$ (Fig.~\ref{fig8}(d)), 
we also show the results for corotating binaries for comparison. 
The figures show that the mass fraction of fluid elements with 
$j/M_{\rm ADM} >1.6$ for irrotational binary neutron stars 
is zero irrespective of $n$ and $\compa$. 
In contrast, that for corotating binaries is $\sim 5\%$ even for 
$j/\madm =2$. 

Now, we assume 
the following hypotheses which are likely to hold in the merger 
of binary neutron stars with $\compa \geq 0.14$: 
(1) A black hole is formed. This assumption could be wrong 
for $n=0.5$ and $\compa \sim 0.14$, 
but would be correct for a moderate stiffness, 
$2/3 \alt n \alt 1$ (for $n=1$, 
numerical simulations have indicated the correctness of this 
assumption \cite{su00}). 
(2) The disk mass formed after the merger 
is much smaller than the black hole mass (say at most 10$\%$ of the 
total rest mass), 
and consequently the mass of black hole $M_{\rm BH}$ 
is nearly equal to the initial gravitational mass $\madm$ (say 
$\geq 0.9\madm$). 
(3) The specific angular momentum $j$ for most of fluid elements 
changes by at most $10 \%$ during the merger process.  
One mechanism to change $j$ is gravitational radiation 
which would decrease $j$ of all fluid elements by $\lo 10 \%$.  
The other is hydrodynamic interaction which could transport 
$j$ outward and consequently a part of fluid elements could have 
larger $j$ than its original value.  Therefore this hypothesis 
would be reasonable if the merger process to form the black hole 
proceeds on the dynamical timescale of the system and it is shorter 
than the timescale for the outward transportation of angular momentum 
by the hydrodynamic interaction \cite{footh3}.  
(4) The quantity $q$ of the black hole finally formed is 
about $10\%$ smaller than the initial value for the system 
because of gravitational radiation reaction 
\cite{footq} ({\it i.e.}, $q \alt 0.9$).

With the hypotheses (1), (2) and (4), the specific angular momentum 
of a test particle at the ISCO around the formed black hole 
is approximately determined from the $M_{\rm ADM}$ and $J_{\rm tot}$ 
given initially.  
As shown above, $q$ for binaries with $\compa \geq 0.14$ is 
less than unity and according to (4), $q$ of the black hole 
will be $\alt 0.9$ after the merger. This implies that 
the specific angular momentum $j$ at the ISCO around the 
black hole would be larger than $2M_{\rm BH}$ \cite{footkb}, 
and to form disks, there have to exist 
a fraction of fluid elements of $j \agt 2 M_{\rm BH} \agt 1.8 \madm$. 
However, as shown in Figs.~\ref{fig8}, the mass of the fluid element 
of $j/M_{\rm ADM}>1.6$ is almost zero before merger for any $n$. 
Thus, if all the hypotheses are correct, 
the mass fraction of the disk around the black hole is almost zero 
after the merger of irrotational binary neutron stars of 
$\compa \geq 0.14$. In contrast, 
the mass fraction of the disk with $j \agt 2 M_{\rm BH}$ 
can be about $\sim 5\%$ for corotating case as shown in 
Fig.~\ref{fig8}(d).

Recent numerical simulations of merging binary neutron stars 
of equal mass in full general relativity for $n=1$ \cite{su00} 
indicate that for irrotational binaries, 
the mass of the disk around the black hole formed after 
merger is negligible, while for corotating binaries, 
it could be $\sim 5-10\%$. These results are in complete 
agreement with the above analysis. This suggests that 
the above working hypotheses are approximately correct. 
Also, it suggests that the present analysis can be applied 
for other $n$. 
Thus, irrespective of $n$ and $\compa(\geq 0.14)$, 
a massive disk is unlikely to be formed around the black hole 
after mergers of irrotational binary neutron stars.

The present result leads us to give a strong caution with 
regard to previous numerical studies.  
As shown above, the fraction of the disk mass depends sensitively 
on the initial velocity field and 
effect of general relativity (which determines the minimum 
allowed value of the specific angular momentum of the disk). 
In the past decade, most of 
simulations of binary neutron stars of equal mass have been performed 
using rather crude initial conditions for the velocity 
field and without general relativistic effects (see, {\it e.g.}, 
\cite{ONS} for review). Such simulations 
in some cases have provided results in which a disk is formed 
around the merged objects. However, the present 
analysis suggests that such results might be wrong 
and gives a warning that it is dangerous to draw any 
conclusion about the mass of the disk 
without fully including general relativistic effects and 
without using correct velocity fields for the initial conditions.

\section{Summary}
\label{sec4}

We have presented numerical results for 
quasiequilibrium sequences of irrotational binary neutron stars 
of equal mass in general relativity adopting the polytropic 
equation of state $p=\kappa \rho^{1+1/n}$. 
Computations have been performed for a wide range of the polytropic 
index $n$ and the compactness $\compa$. 
We have analyzed the sequences from various points of view. 
The conclusions obtained in this paper are as follows.

The dynamical instability for orbits of irrotational binary 
neutron stars sets in before 
mass transfer occurs for $n \lo 2/3$ $(\Gamma \go 2.5)$.
Thus, the binary reaches the ISCO without mass transfer in this case. 
On the other hand, for $n \go 2/3$ $(\Gamma \lo 2.5)$, 
mass transfer will occur before binaries reach the ISCO.  
This property depends very weakly 
on the compactness for $\compa \lo 0.17$.  
The frequency of gravitational waves 
at the innermost orbit (ISCO or the point where mass transfer 
sets in) is about 800 to 1500 Hz 
depending on the compactness.  For a binary of 
less massive ({\it i.e.}, less compact) neutron stars, the frequency 
could be less than 1000 Hz and as a result 
the characteristic signal of gravitational waves 
emitted around the ISCO may be detected by laser 
interferometric detectors.

We have determined the ISCOs for a wide range of $\compa \leq 0.3$ 
for $n=0.5$ and of $0.17 \leq \compa \leq 0.27$ for $n=0.66667$. 
The results indicate that 
the ISCO is determined by the hydrodynamic instability and 
not by the general relativistic orbital instability for 
realistic binary neutron stars with $0.14 \alt \compa \alt 0.2$. 
Our present results indicate that the approaches using 
equations of motion for point particles 
are not appropriate for determining 
the ISCO of binary neutron stars if the neutron stars 
are not extremely compact $\compa \gg 0.2$.

The maximum density of neutron stars in irrotational 
binary systems decreases as the orbital separation decreases.  
Associated with this, we show that neutron stars in irrotational 
binary systems (which are stable in isolation) 
are stable against gravitational collapse until the merger 
sets in (see Appendix for details).

The angular momentum parameter $q=\jom$ is less than unity for 
$\compa \agt 0.13$ irrespective of 
$n$ for irrotational binaries at the innermost orbits 
({\it i.e.} $\bar d_G=\bar d_R$ or $\bar d_{\rm dyn}$). 
Since $\compa$ of realistic neutron stars is expected to be larger 
than $0.14$, 
the result implies that realistic binary neutron stars satisfy 
a necessary condition for forming black hole, $q < 1$, 
before the merger. However, this condition is not sufficient for 
determining the outcome after the merger as mentioned in 
Sec.~\ref{sec3e}.  
Numerical simulations in full general 
relativity are obviously necessary to clarify the nature of 
the object resulting from the merger \cite{su00}.

It is found that the specific angular momentum of 
all of the fluid elements in irrotational binary 
neutron stars of equal mass will become very small ($j/\madm < 1.6$) 
before merger irrespective of $n$ and $\compa$. 
As long as outward transportation of angular momentum 
does not work effectively during merging, 
the disk mass around a black hole which would be 
formed after the merger for $\compa \agt 0.14$ will be negligible.  
We should note here that the present conclusion is obtained for 
binaries of equal mass (or nearly equal mass). In the case 
when the mass ratio of two stars deviates from unity by a large factor, 
the star of smaller mass could be 
tidally disrupted by the companion of larger mass 
at a fairly large orbital separation at which 
the specific angular momentum of fluid elements could be 
large enough to form a disk during the tidal disruption. 
For exploring the possibility for formation of a disk in this way, 
it is necessary to perform computations for binary neutron stars of 
unequal mass, and this is one of the key issues for the future work.

\acknowledgments

We would like to thank Prof. J. C. Miller for 
carefully reading of the manuscript, helpful comments and 
continuous encouragement.  We also would like to thank Dr. M. 
Ruffert for useful discussions and comments on the manuscript.  
K.U. would like to thank Prof. D. W. Sciama for his warm 
hospitality at SISSA and ICTP.  M.S. thanks JSPS and the 
Department of Physics, University of Illinois for hospitality.  
Numerical computations were in part carried out at the 
Astronomical Data Analysis Center of the National Astronomical 
Observatory of Japan. 

\appendix

\section*{Stability of neutron stars against 
gravitational collapse and 
existence of supramassive irrotational 
binary neutron stars}\label{calib}

It is important to check that neutron stars which are stable in isolation 
are stable against gravitational collapse to black holes even in 
binary systems.  
(See \cite{wm956,mmr99} about a claim that the neutron stars 
in irrotational binary systems may be unstable to collapse individually.)  
In this Appendix, we show that no evidence is found in our numerical 
experiment for selected values of $n$. 

The concern about gravitational collapse is 
relevant for neutron stars of near to the maximum allowed rest mass. 
Thus we here focus only on such massive neutron stars. 
As noted in the caption of Table I, the compactness $\compa$ of 
neutron stars at the maximum allowed rest mass is 
larger than $0.25$ for $n\leq 0.8$, $\sim 0.21$ for $n=1$ and 
$\sim 0.17$ for $n=1.25$. 
The compactness of realistic neutron stars 
would be in a range between $0.14 \alt \compa \alt 0.2$ 
and probably less than 0.25. 
{}From this fact, we pay attention only to $n=1$ and 1.25 
here.

First, for calibrating our numerical code, 
we compare numerical results at large $\whd$ with those for 
spherical stars computed by solving the TOV equations. 
Then, we show the existence of supramassive neutron stars in 
close binary systems confirming 
the excess of the maximum rest mass beyond the maximum rest mass of 
the spherical stars. 
Finally, we discuss the stability against collapse of 
supramassive neutron stars in close binary systems 
(stars whose rest mass is larger than the spherical maximum mass).

In the analysis, we computed quasiequilibrium states with 
fixed $\whd$, changing $\compa$ (or $\brmx$).  
In Fig.~\ref{fig9}, we show $\bar M_0$ as a function of $\brmx$ 
(solid lines)  (a) for $n=1$ and (b) for $n=1.25$. Here, 
$\whd$ is chosen between $1.3125$ and $2.0$ for $n=1$ and 
between $1.3125$ and $2.5$ for $n=1.25$. 
For obtaining the configuration at $\whd = \infty$ (denoted by 
the dashed lines), the TOV equations are solved.  
Note that the maximum mass for spherical stars is 
$\bMo \simeq 0.180$ at $\brmx \simeq 0.318$ ($\compa \simeq 0.214$) 
for $n=1$ and 
$\bMo \simeq 0.2167$ at $\brmx \simeq 0.148$ ($\compa \simeq 0.172$) 
for n=1.25.

For large separation $\whd \go 2.0$, neutron stars in 
irrotational binary systems become almost spherical irrespective of $n$. 
Accordingly, the solid curves should converge toward 
the dashed curve with increasing $\whd$. 
Figure~\ref{fig9} shows that this is approximately the case, 
but the curve in the limit $\whd \gg 1.0$  
slightly deviates from the dashed curve.  
This deviation is due to a systematic error in numerical computation.  
This shows that the present numerical code overestimates the rest mass by 
$\sim 0.4 \%$ for a quasiequilibrium state with a given $\brmx$.  
{}For this reason, the curve in the limit $\whd \gg 1.0$ 
denotes the relation 
between the rest mass and $\brmx$ for spherical stars 
{\it computed in our numerical code}, and hereafter, we 
refer to this curve as the relation for spherical stars. 
The maximum of this curve is 
$\bMo^{\rm smax}$ $=0.1802$ for $n=1$ and $=0.2174$ for $n=1.25$. 

With decreasing $\whd$, the maximum value of 
$\bMo(\brmx)$ increases. 
This suggests that the maximum allowed rest mass of neutron stars in 
irrotational binary systems is increased due to the tidal effect.  
The maximum at $\whd=1.3125$ 
is $\bMo \simeq 0.1808$ for $n=1$ and $0.2179$ for $n=1.25$. 
This implies that supramassive neutron stars exist in 
irrotational binary systems, but 
the maximum mass $\bMo$ can be increased by at most $\sim 0.3\%$ 
even at $\bdg \simeq \bdR$. This increase is much 
smaller than that in the corotating binaries \cite{ba97}, 
for which $\bMo$ increases by $\sim 2 \%$ for $n=1$ and 
and by $\sim 1.5 \%$ for $n=1.5$.  
The reason is that neutron stars in irrotational binary systems have 
negligible spin, while those in corotating binary systems have 
significant spin and the centrifugal force by this can increase 
the maximum allowed rest mass.  Such binary systems with 
supramassive neutron stars could be formed in principle after 
a supramassive core collapse to undergo bifurcation.

Baumgarte et al. \cite{ba97} have discussed the stability of 
a supramassive neutron star in a {\it corotating} binary 
using the turning point method \cite{turning}. 
They have concluded that the supramassive 
and all other neutron stars below a critical density 
in a corotating binary are stable against gravitational collapse.  
Following their discussion, 
we plot in Fig.~\ref{fig10} sequences of constant $\bMo$ for 
a variety of $\bMo$ on the $\brmx$--$\bM$ plane 
(the solid and thick long dashed lines).  
Each solid curve is drawn for $1.3125 \leq \whd \leq 2.0$ in the case
$n=1$ and for $1.3125 \leq \whd \leq 1.875$ in the case $n=1.25$. 
Here, each sequence of constant $\bMo$ can be 
interpreted as an evolutionary sequence 
of a binary system as discussed in section \ref{sec3}.  
When a binary system evolves as a result of 
gravitational radiation, the gravitational mass decreases.
Thus, if $\brmx$ decreases (increases) with decreasing 
gravitational mass, the neutron stars are expected 
to be in a stable (unstable) branch.  
Therefore we may conclude that the curves in region 
(i) are stable while those in region (iii) are unstable.

For the thick dashed curves in region (ii), a turning point exists at 
a maximum $\bM$.  In contrast to the corotating binary case, 
it is not fully established whether the stability of a solution 
sequence changes at this turning point or not.  
However, from the tendency of solution the sequences (i), (ii) 
and (iii), it is reasonable to expect that a part of the solutions 
on the left branches of the thick dashed curves in region (ii) 
are stable against gravitational collapse.  
{}From these results, it is reasonable to conclude that stable 
supramassive neutron stars can exist in irrotational binary systems 
as a result of tidal deformation, although the excess of the 
rest mass beyond the maximum rest mass of the spherical stars 
is tiny.

\clearpage

\begin{table*}
\begin{center}
\begin{tabular}{ccccccccccc}
\\[-3mm]
& $\compa$ & $\whd$ & $\brmx$ & $\bMo$ & $\bM$ & $M_0 \Omega $ & $\jom$ & 
$\dg/M$ & \\[1mm]
\tableline
\\[-3mm]
\multicolumn{10}{c}{$n=0.5$} \\[1mm]
\tableline
\\[-2.5mm]
& 0.10 & 1.1250 & 2.50E-01 & 4.41E-02 & 4.12E-02 & 1.07E-02 & 1.12E+00 & 1.28E+01 & \\[1mm]
& 0.12 & 1.1250 & 2.83E-01 & 5.52E-02 & 5.08E-02 & 1.46E-02 & 1.05E+00 & 1.03E+01 & \\[1mm]
& 0.14 & 1.1250 & 3.16E-01 & 6.65E-02 & 6.04E-02 & 1.92E-02 & 9.95E-01 & 8.56E+00 & \\[1mm]
& 0.17 & 1.1250 & 3.69E-01 & 8.40E-02 & 7.45E-02 & 2.72E-02 & 9.41E-01 & 6.67E+00 & \\[1mm]
& 0.19 & 1.1250 & 4.08E-01 & 9.56E-02 & 8.36E-02 & 3.35E-02 & 9.15E-01 & 5.74E+00 & \\[1mm]
\tableline
\\[-3mm]
\multicolumn{10}{c}{$n=0.66667$} \\[1mm]
\tableline
\\[-2.5mm]
& 0.10 & 1.1875 & 1.60E-01 & 6.06E-02 & 5.68E-02 & 1.06E-02 & 1.10E+00 & 1.28E+01 & \\[1mm]
& 0.12 & 1.1875 & 1.90E-01 & 7.41E-02 & 6.84E-02 & 1.45E-02 & 1.03E+00 & 1.03E+01 & \\[1mm]
& 0.14 & 1.1875 & 2.21E-01 & 8.74E-02 & 7.97E-02 & 1.89E-02 & 9.81E-01 & 8.56E+00 & \\[1mm]
& 0.17 & 1.1875 & 2.75E-01 & 1.07E-01 & 9.55E-02 & 2.68E-02 & 9.28E-01 & 6.69E+00 & \\[1mm]
& 0.19 & 1.1875 & 3.16E-01 & 1.19E-01 & 1.05E-01 & 3.30E-02 & 9.03E-01 & 5.73E+00 & \\[1mm]
\tableline
\\[-3mm]
\multicolumn{10}{c}{$n=0.8$} \\[1mm]
\tableline
\\[-2.5mm]
& 0.10 & 1.1875 & 1.13E-01 & 7.77E-02 & 7.30E-02 & 1.07E-02 & 1.10E+00 & 1.27E+01 & \\[1mm]
& 0.12 & 1.1875 & 1.39E-01 & 9.32E-02 & 8.64E-02 & 1.45E-02 & 1.03E+00 & 1.03E+01 & \\[1mm]
& 0.14 & 1.1875 & 1.69E-01 & 1.08E-01 & 9.88E-02 & 1.89E-02 & 9.75E-01 & 8.52E+00 & \\[1mm]
& 0.17 & 1.1875 & 2.21E-01 & 1.28E-01 & 1.15E-01 & 2.68E-02 & 9.22E-01 & 6.63E+00 & \\[1mm]
& 0.19 & 1.1875 & 2.63E-01 & 1.41E-01 & 1.25E-01 & 3.28E-02 & 8.97E-01 & 5.71E+00 & \\[1mm]
\tableline
\\[-3mm]
\multicolumn{10}{c}{$n=1.0$} \\[1mm] 
\tableline
\\[-2.5mm]
& 0.10 & 1.2500 & 6.96E-02 & 1.12E-01 & 1.05E-01 & 1.06E-02 & 1.09E+00 & 1.27E+01 & \\[1mm]
& 0.12 & 1.2500 & 9.12E-02 & 1.30E-01 & 1.21E-01 & 1.44E-02 & 1.02E+00 & 1.03E+01 & \\[1mm]
& 0.14 & 1.2500 & 1.17E-01 & 1.46E-01 & 1.35E-01 & 1.87E-02 & 9.71E-01 & 8.52E+00 & \\[1mm]
& 0.17 & 1.2500 & 1.69E-01 & 1.66E-01 & 1.50E-01 & 2.63E-02 & 9.19E-01 & 6.66E+00 & \\[1mm]
& 0.19 & 1.2500 & 2.15E-01 & 1.75E-01 & 1.57E-01 & 3.20E-02 & 8.95E-01 & 5.75E+00 & \\[1mm]
\tableline
\\[-3mm]
\multicolumn{10}{c}{$n=1.25$} \\[1mm]
\tableline
\\[-2.5mm]
& 0.10 & 1.2500 & 3.96E-02 & 1.73E-01 & 1.64E-01 & 1.06E-02 & 1.08E+00 & 1.26E+01 & \\[1mm]
& 0.12 & 1.2500 & 5.65E-02 & 1.92E-01 & 1.80E-01 & 1.43E-02 & 1.02E+00 & 1.02E+01 & \\[1mm]
& 0.14 & 1.2500 & 7.93E-02 & 2.07E-01 & 1.92E-01 & 1.85E-02 & 9.67E-01 & 8.50E+00 & \\[1mm]
& 0.17 & 1.2500 & 1.19E-01 & 2.17E-01 & 2.00E-01 & 2.40E-02 & 9.25E-01 & 7.02E+00 & \\[1mm]
\end{tabular}
\end{center}
\caption{
Characteristic quantities for irrotational binary neutron stars 
for configurations at $\bdg = \bdR$.  Here, $\bar \rho_{\rm max}$ 
is the maximum rest mass density. 
Note that for spherical stars, the maximum value of the normalized 
rest mass and the compactness at the maximum $(\bar M_0, \compa)$ are 
(0.152, 0.316) for $n=0.5$, (0.155, 0.278) for n=0.66667, 
(0.162, 0.252) for $n=0.8$, (0.180, 0.214) for n=1, and 
(0.217, 0.172) for n=1.25. 
Because of an uncertainty for determination of the configuration 
with cusps, $M_0\Omega$ and $\madm \Omega$ 
have an error of $1-2\%$ for each $\compa$. 
\label{tabjm}}
\end{table*}
\clearpage

\begin{table*}
\begin{center}
\begin{tabular}{ccccccccccc}
\\[-3mm]
& $\compa$ & $\whd$ & $\brmx $ & $\bMo$ & $\bM$ & $M_0 \Omega$ & 
$\madm \Omega$ & $\jom$ & $\dg/M$ & \\[1mm]
\tableline
\\[-2.5mm]
& 0.03 &1.1875 & 0.124 & 9.89E-03 & 9.69E-03 & 1.55E-03 & 3.04E-03 & 1.88 & 47.7 & \\[1mm]
& 0.06 &1.1875 & 0.183 & 2.34E-02 & 2.25E-02 & 4.59E-03 & 8.83E-03 & 1.38 & 22.9 & \\[1mm]
& 0.10 &1.2500 & 0.251 & 4.41E-02 & 4.12E-02 & 1.04E-02 & 1.94E-02 & 1.12 & 13.1 & \\[1mm]
& 0.12 &1.2500 & 0.284 & 5.52E-02 & 5.08E-02 & 1.41E-02 & 2.60E-02 & 1.05 & 10.6 & \\[1mm]
& 0.14 &1.2500 & 0.317 & 6.65E-02 & 6.04E-02 & 1.85E-02 & 3.36E-02 & 0.994 & 8.76 & \\[1mm]
& 0.17 &1.3125 & 0.371 & 8.40E-02 & 7.45E-02 & 2.55E-02 & 4.53E-02 & 0.938 & 6.97 & \\[1mm]
& 0.19 &1.3125 & 0.410 & 9.56E-02 & 8.36E-02 & 3.14E-02 & 5.50E-02 & 0.912 & 6.00 & \\[1mm]
& 0.21 &1.3125 & 0.457 & 1.07E-01 & 9.22E-02 & 3.84E-02 & 6.60E-02 & 0.899 & 5.20 & \\[1mm]
& 0.24 &1.3750 & 0.529 & 1.24E-01 & 1.04E-01 & 4.83E-02 & 8.13E-02 & 0.872 & 4.37 & \\[1mm]
& 0.27 &1.5000 & 0.628 & 1.39E-01 & 1.14E-01 & 5.79E-02 & 9.54E-02 & 0.858 & 3.81 & \\[1mm]
& 0.30 &1.6250 & 0.773 & 1.49E-01 & 1.21E-01 & 6.79E-02 & 0.110    & 0.851 & 3.37 & \\[1mm]
\end{tabular}
\end{center}
\caption{
The same as Table I but for a configuration with $n=0.5$ 
at $\bdg = \bddy$. We include $\madm \Omega$ in this Table. 
Because of an uncertainty for determining 
the minima of $J_{\rm tot}$ and $\madm$, $M_0\Omega$ and $\madm \Omega$ 
have an error of $1-2\%$ for each $\compa$. 
\label{tabjmn05}}
\end{table*}
\begin{table*}
\begin{center}
\begin{tabular}{ccccccccccc}
\\[-3mm]
& $\compa$ & $\whd$ & $\brmx $ & $\bMo$ & $\bM$ & $M_0 \Omega$ 
& $\madm \Omega$ & $\jom$ & $\dg/M$ & \\[1mm]
\tableline
\\[-2.5mm]
& 0.17 &1.1875 & 2.75E-01 & 1.07E-01 & 9.55E-02 & 2.68E-02 & 4.78E-02 & 0.928 & 6.69 & \\[1mm]
& 0.19 &1.1875 & 3.16E-01 & 1.19E-01 & 1.05E-01 & 3.30E-02 & 5.82E-02 & 0.903 & 5.73 & \\[1mm]
& 0.21 &1.2500 & 3.64E-01 & 1.31E-01 & 1.14E-01 & 3.91E-02 & 6.79E-02 & 0.884 & 5.06 & \\[1mm]
& 0.24 &1.3125 & 4.56E-01 & 1.45E-01 & 1.24E-01 & 4.93E-02 & 8.42E-02 & 0.865 & 4.24 & \\[1mm]
& 0.27 &1.4375 & 5.85E-01 & 1.54E-01 & 1.30E-01 & 5.78E-02 & 9.75E-02 & 0.855 & 3.74 &
\end{tabular}
\end{center}
\caption{
The same as Table II but for a configuration with $n=0.66667$ 
at $\bdg = \bddy$.  Because of an uncertainty for determining 
the minima of $J_{\rm tot}$ and $\madm$, $M_0\Omega$ has an error 
of $1-2\%$ for each $\compa$. 
Note that $\bddy$ is nearly equal to $\bdR$ for $\compa \leq 0.19$ 
but not for $\compa > 0.2$.  
\label{tabjmn66}}
\end{table*}
\clearpage
\begin{table}
\begin{center}
\begin{tabular}{cccc}
\\[-3mm]
& $\compa$ & $(a,\ b,\ c)$ & \\[1mm]
\tableline
\\[-2.5mm]
&  $0.03 - 0.17$ & $(0.2849, 0.422, 0.29)$ & \\[1mm]
&  $0.03 - 0.19$ & $(0.2878, 0.392, 0.93)$ & \\[1mm]
\tableline
\\[-2.5mm]
&  $\compa$ & $(a',\ b',\ c')$ & \\[1mm]
\tableline
\\[-2.5mm]
&  $0.03 - 0.17$ & $(0.2849, 0.245,-0.08)$ & \\[1mm]
&  $0.03 - 0.19$ & $(0.2864, 0.200, 0.18)$ & \\[1mm]
\end{tabular}
\end{center}
\caption{
Coefficients of the fitting formula Eq.~(\ref{fiteq}) 
for the results for $n=0.5$ (Table II).  The column 
headed $\compa$ shows the data sets used for each fitting. 
\label{tab4}
}
\end{table}
\vfill
%
\vfill
\vfill
%

%
\vfill
\vfill
\vfill
\vfill
%
%
%
%
\vfill
%
%
%
\begin{figure}
\epsfxsize=8truecm
\begin{center}
\epsffile{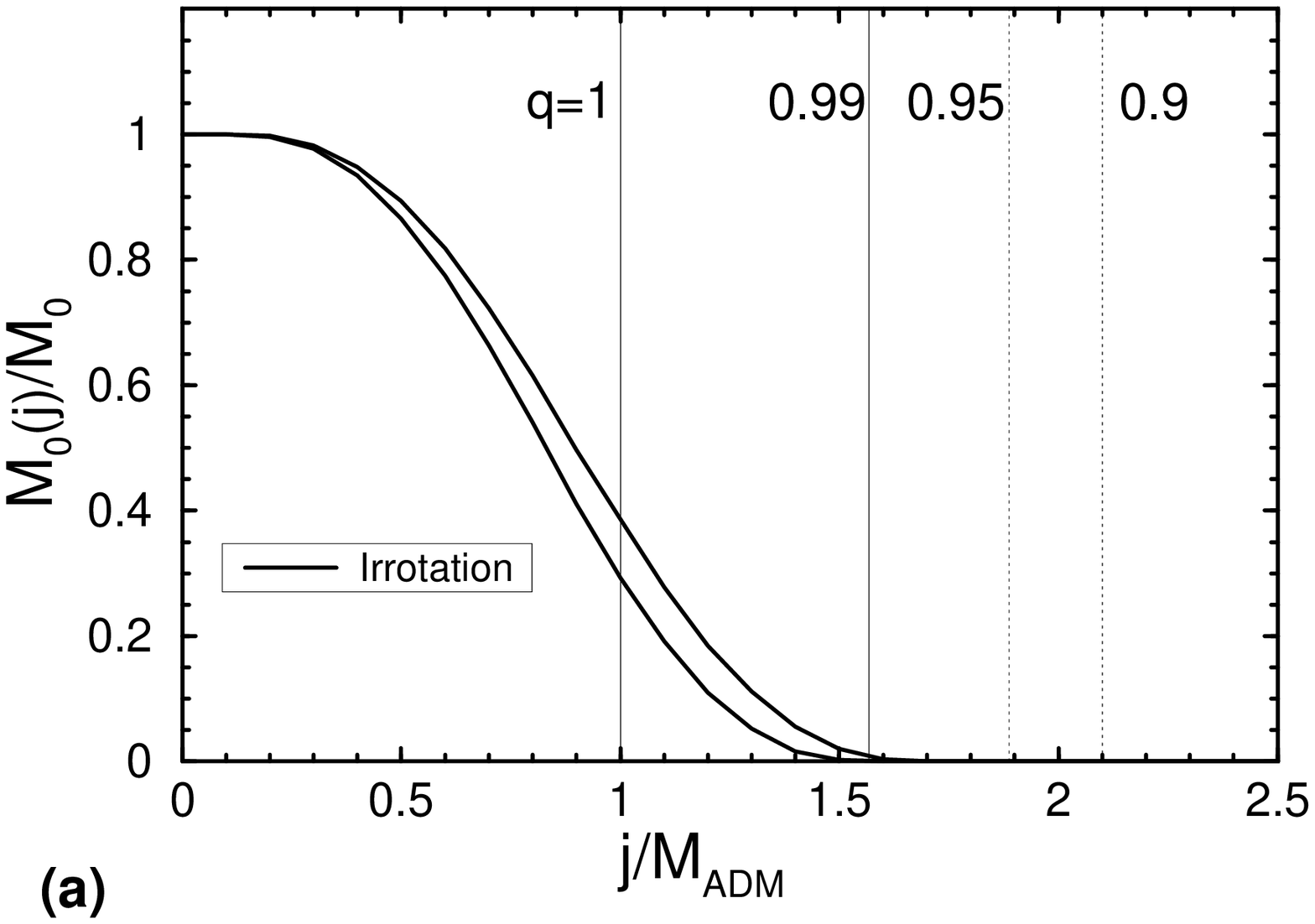}
\end{center}
\epsfxsize=8truecm
\begin{center}
\epsffile{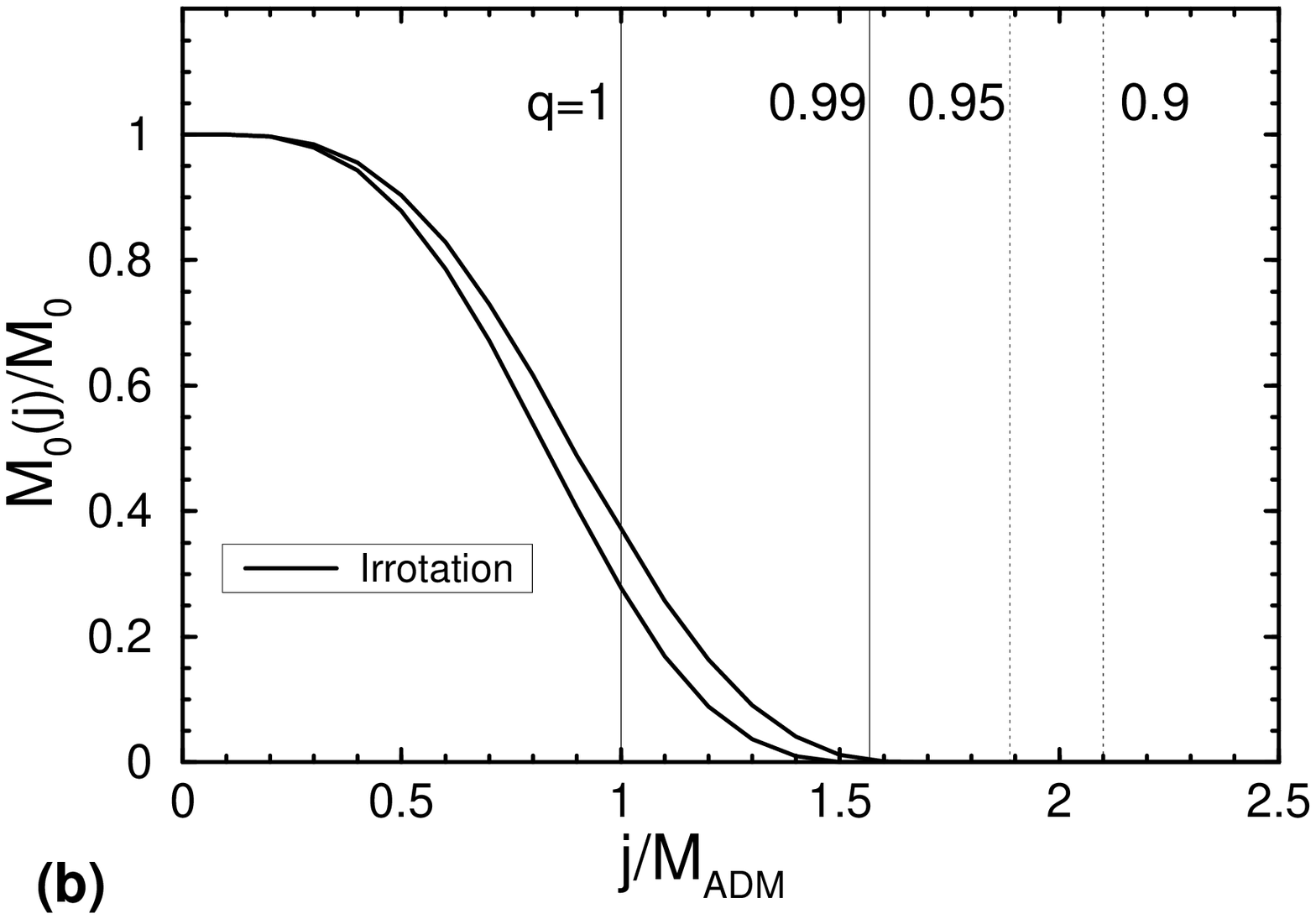}
\end{center}
\epsfxsize=8truecm
\begin{center}
\epsffile{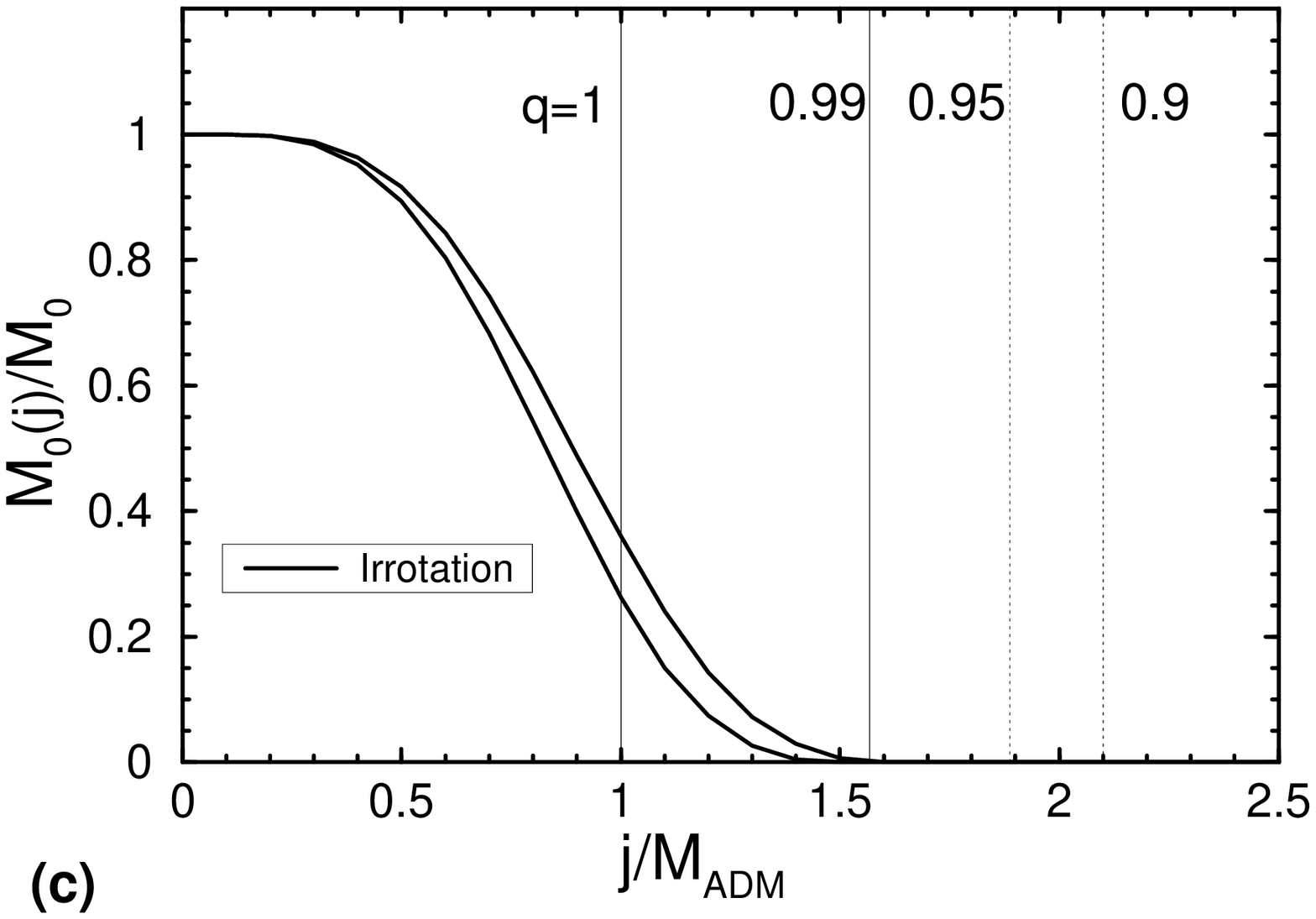}
\end{center}
\end{figure}
\vfill
%
%
\begin{figure}
\epsfxsize=8truecm
\begin{center}
\epsffile{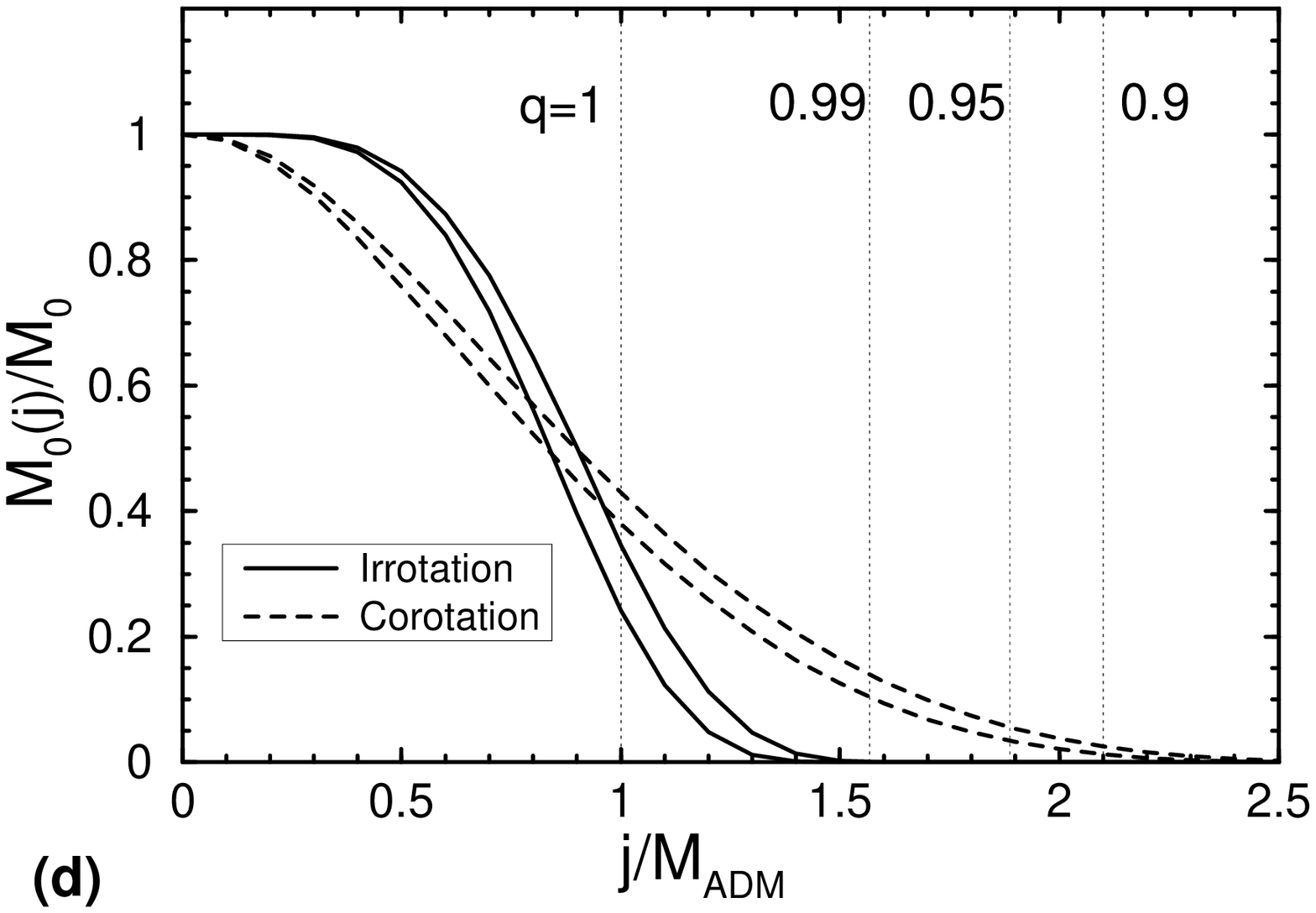}
\end{center}
\epsfxsize=8truecm
\begin{center}
\epsffile{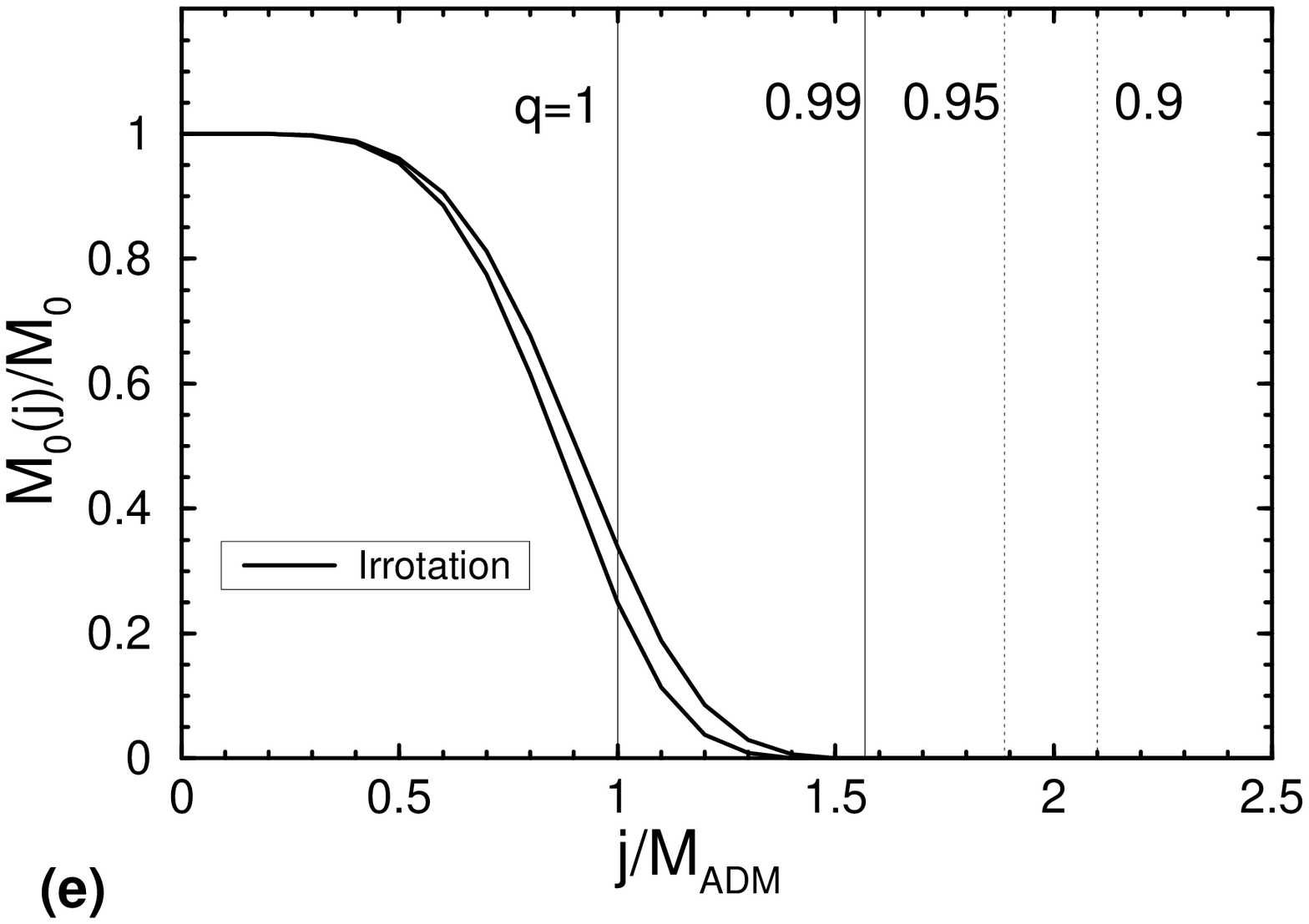}
\end{center}
\caption{
Fraction of rest mass with specific angular momentum 
larger than $j$ for the innermost orbits 
for $\compa=0.14$ and $0.17$ and for 
(a) $n=0.5$, (b) $n=0.66667$, (c) $n=0.8$, 
(d) $n=1$, and (e) $n=1.25$. The 
solid and dashed lines correspond to the results for 
irrotational and corotating binaries, respectively. 
The lines for $\compa=0.17$ appear as upper lines in each panel.  
For the irrotational case, the models at $\bdg=\bddy$ are chosen 
for $n = 0.5$, and the models at $\bar d_G=\bar d_R$ are chosen 
for the other $n$.  For the corotating case, the models for which 
the two surfaces come into contact are chosen. 
\label{fig8}}
\end{figure}
\vfill
%
%
%
%
%
\begin{figure}
\epsfxsize=8truecm
\begin{center}
\epsffile{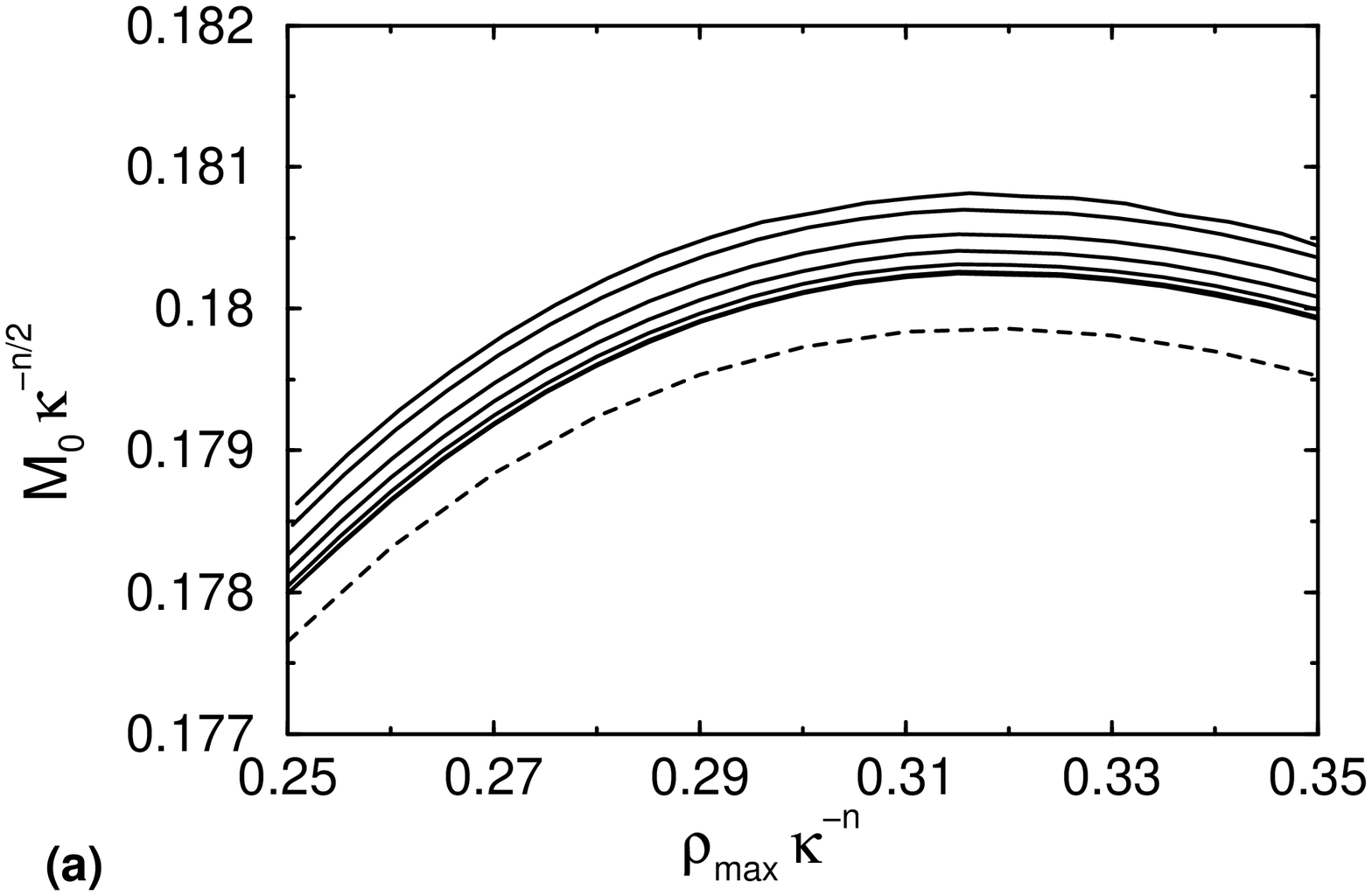}
\end{center}
\epsfxsize=8truecm
\begin{center}
\epsffile{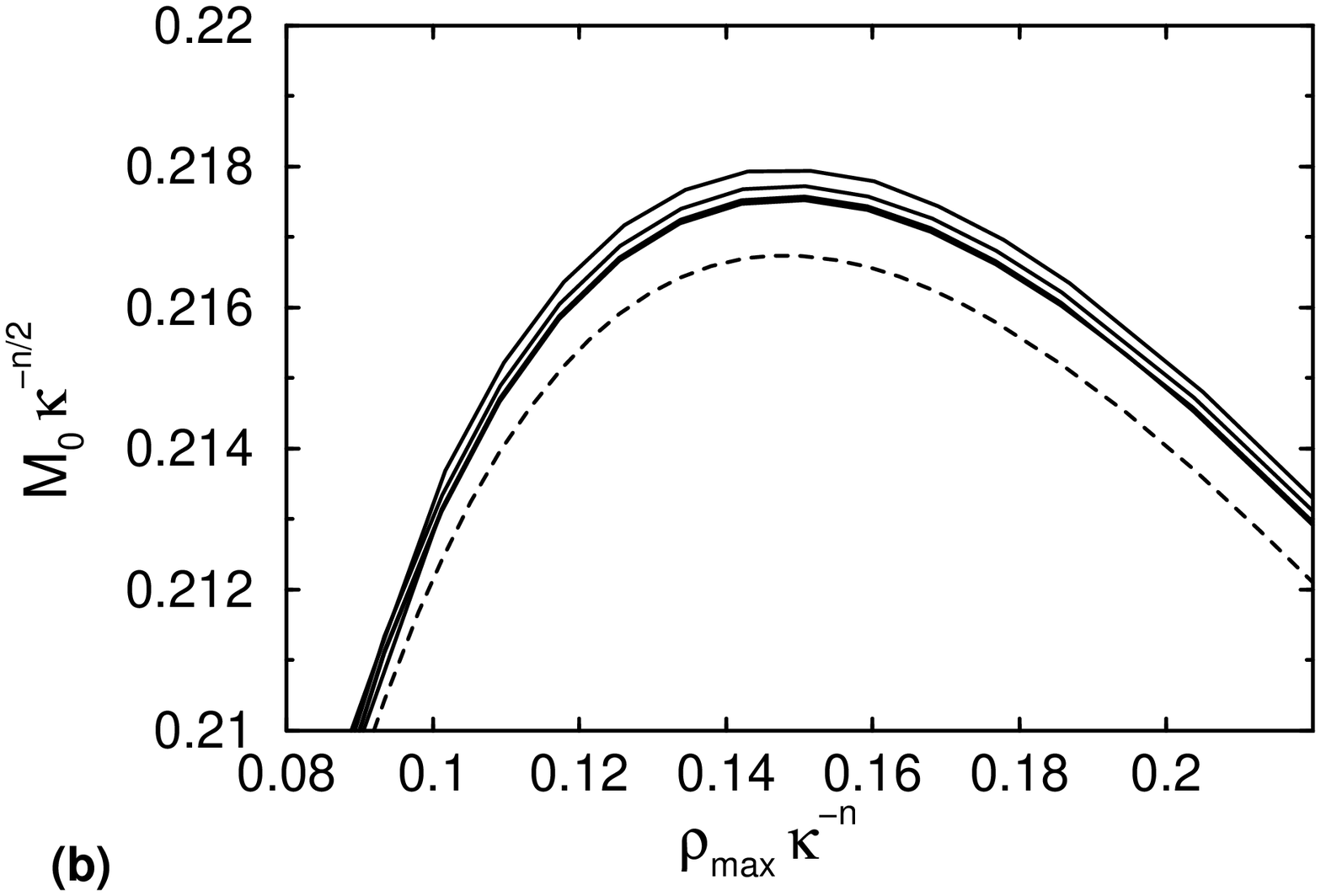}
\end{center}
\caption{
$\bMo$ as a function of $\brmx$
(a) Solid lines from top to bottom correspond to 
$\whd = 1.3125$, 1.375, 1.5, 1.625, 1,75, 1,875 and 2 
for the $n=1$ case.  
(b) Solid lines from top to bottom correspond to 
$\whd = 1.3125, 1.5, 1.75, 2$ and $2.5$ for the $n=1.25$ case.  
A dashed line corresponds to a solution of the TOV equations in 
each panel.  
\label{fig9}}
\end{figure}
\vfill
%
%
%
\begin{figure}
\epsfxsize=8truecm
\begin{center}
\epsffile{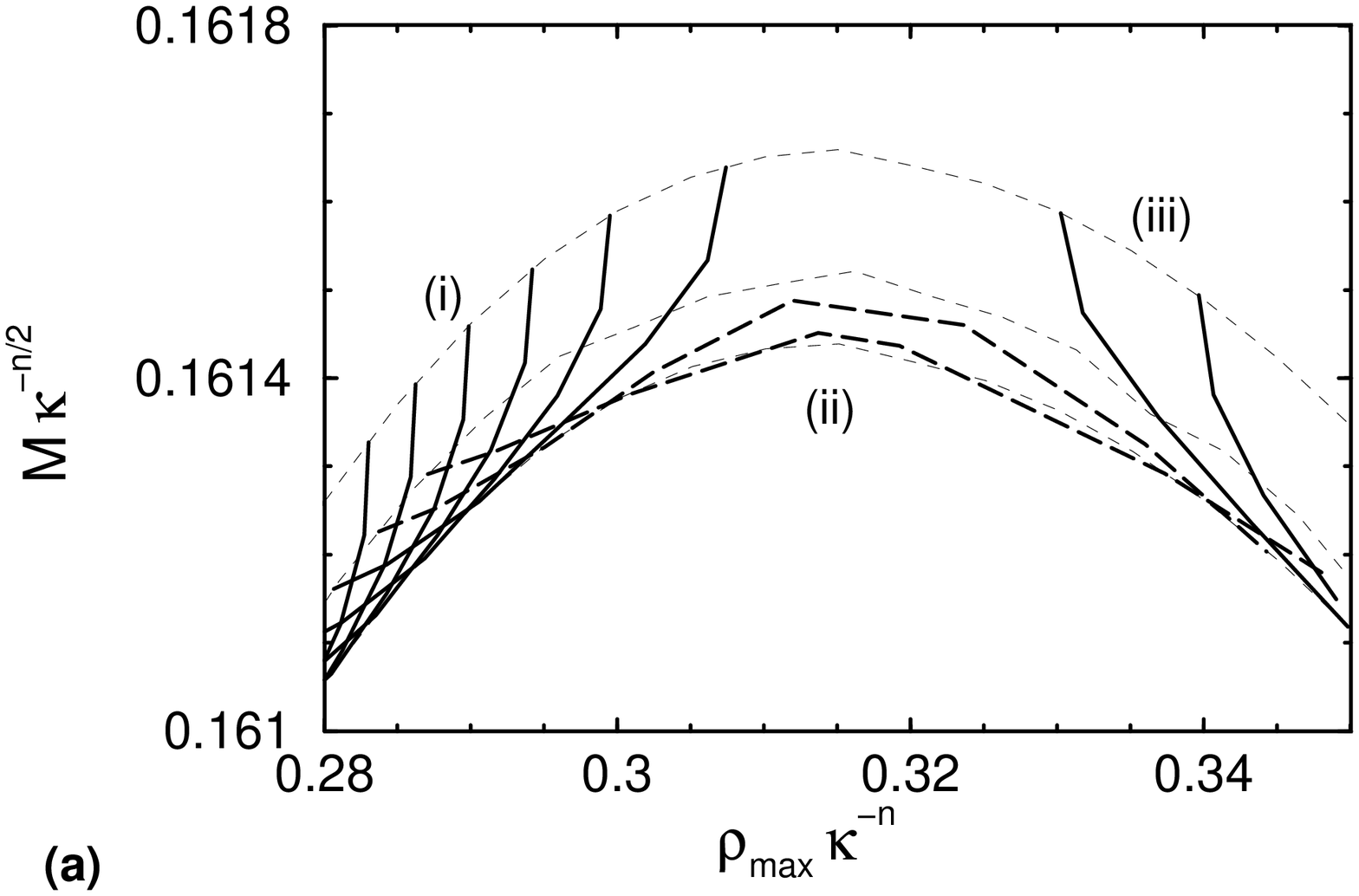}
\end{center}
\epsfxsize=8truecm
\begin{center}
\epsffile{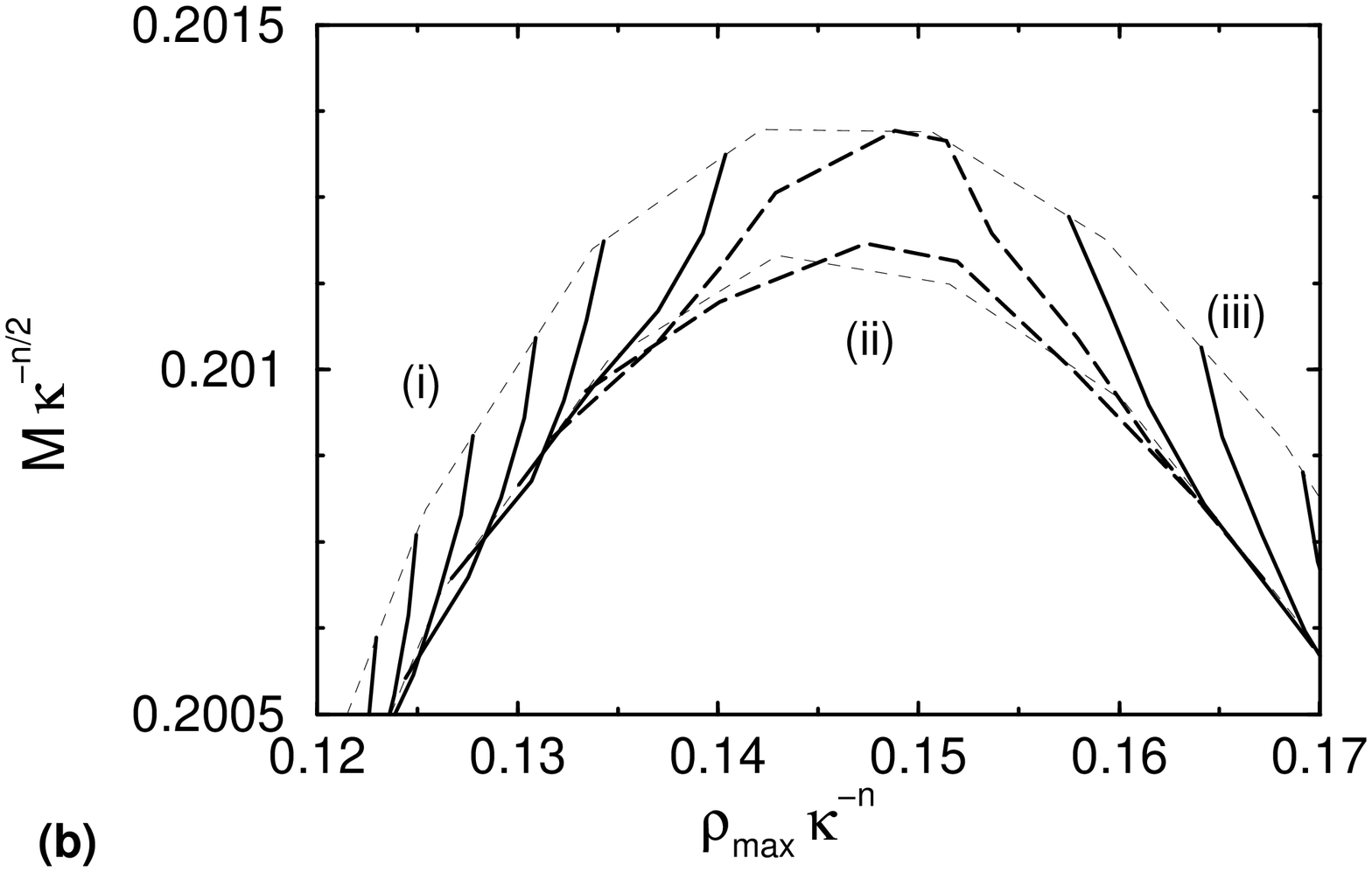}
\end{center}
\caption{
Sequences for constant $\bMo$ are plotted in the $\brmx-\bM$ 
plane (thick solid lines and thick long dashed lines).  
(a) $n=1$ case.  
{}From left to right, the $\bMo$ of each curve 
changes from $0.1797$ to $0.1802$ in steps of $0.0001$ 
(solid lines) in region (i) ({\it i.e.}, $\brmx < 0.31$). 
In region (ii), two long dashed 
lines with $\bMo = 0.1803$ and $0.1804$ (upper one and lower one 
respectively) are plotted. In region (iii) ({\it i.e.}, $\brmx > 0.31$), 
$\bMo$ is taken from $0.1802$ to $0.1801$ 
in steps of $0.0001$ (solid lines).  
The upper, middle and lower thin dashed curves correspond to the 
relation between $\bM$ and $\brmx$ 
for $\whd = 2.0, 1.3125$ and $1.5$, respectively.  
(b) $n=1.25$ case.  
{}From left to right, the $\bMo$ of each curve 
changes from $0.2164$ to $0.2174$ in steps of $0.0002$ 
(solid lines) in region (i) ({\it i.e.}, $\brmx < 0.14$). 
In region (ii), two long dashed 
lines with $\bMo = 0.2175$ and $0.2176$ (upper one and lower one 
respectively) are plotted. In region (iii) ({\it i.e.}, $\brmx > 0.14$), 
$\bMo$ is taken from $0.2174$ to $0.2170$ 
in steps of $0.0002$ (solid lines).  
The upper and lower thin dashed curves correspond to the 
relations between $\bM$ and $\brmx$ 
for $\whd = 1.875$ and $1.3125$, respectively.  
\label{fig10}
}
\end{figure}


\begin{thebibliography}{99}
%
\bibitem{cutler} C. Cutler et al., Phys. Rev. Lett. {\bf 70}, 1984 
(1993); \\
K. S. Thorne, in {\it Proceeding of Snowmass 
95 Summer Study on Particle and Nuclear Astrophysics and 
Cosmology}, edited by E. W. Kolb and R. Peccei (World Scientific, 
Singapore, 1995), p. 398. 

\bibitem{gwref} A. Abramovici et al. Science {\bf 256}, 325 (1992) ; 
C. Bradaschia, et al. Nucl. Instrum. and Methods {\bf A289}, 518 (1990) ; 
J. Hough, in {\it Proceedings of the Sixth Marcel Grossmann Meeting},
edited by H. Sato and T. Nakamura
(World Scientific, Singapore, 1992), p.192 ; 
K. Kuroda et al., in {\it Proceedings of the international conference on
gravitational waves: Sources and Detectors}, edited by
I. Ciufolini and F. Fidecard(World Scientific, 1997), p.100.

\bibitem{piran} {\it e.g.}, T. Piran, in 
{\it Unsolved Problems in Astrophysics},  
edited by J. N. Bahcall and J. P. Ostriker 
(Princeton University Press, 1997), 343.

\bibitem{LP} L.-X. Li and B. Paczynski, Astrophys. J. Lett. {\bf 507}, 
L59 (1998). 

\bibitem{blanchet} {\it e.g.}, L. Blanchet, in {\it 
Relativistic Gravitation and Gravitational Radiation} edited by 
J.-P. Lasota and J.-A. Marck (Cambridge University Press, Cambridge, 
England, 1997), 33; Prog. Theor. Phys. Suppl. {\bf 136}, 146 (1999): \\
C. M. Will, Prog. Theor. Phys. Suppl. {\bf 136}, 158 (1999).

\bibitem{centrella} X. Zhuge, J. M. Centrella, and S. L. W. McMillan, 
Phys. Rev. D {\bf 50}, 6247 (1994); {\bf 54}, 7261 (1996). 

\bibitem{footligo} 
Note however that if a dual recycling technique for a 
narrow band detector is available, it may be still 
possible to detect gravitational waves of high frequency beyond 
$\sim 1$kHz \cite{cutler}. 


\bibitem{on97}
K. Oohara and T. Nakamura, in {\it Relativistic gravitation and gravitational 
radiation}, edited by J.-P. Lasota and J.-A. Marck (Cambridge University 
Press, Cambridge, 199), 309; 
K. Oohara and T. Nakamura, Prog. Theor. Phys. Suppl. {\bf 136}, 270 (1999).  

\bibitem{ss99}
J. A. Font, M. Miller, W.-M. Suen, and M. Tobias,  Phys. Rev. D {\bf 61}, 
044011 (2000); W.-M. Suen, Prog. Theor. Phys. Suppl. {\bf 136}, 251 (1999). 

\bibitem{sh99}
M. Shibata, Prog. Theor. Phys. {\bf 101}, 251 (1999); 
M. Shibata, Prog. Theor. Phys. {\bf 101}, 1199 (1999); 
M. Shibata, Phys. Rev. D {\bf 60}, 104052 (1999). 

\bibitem{su00}
M. Shibata and K. Ury\=u, Phys. Rev. D {\bf 61}, 064001 (2000). 

\bibitem{lt99}
W. Landry and S. A. Teukolsky, preprint gr-qc/9912004 (1999). 

\bibitem{kbc92}
C. S. Kochanek, Astrophys. J. {\bf 398}, 234 (1992);
L. Bildsten and C. Cutler, Astrophys. J. {\bf 400}, 175 (1992).

\bibitem{bas978}
S. Bonazzola, E. Gourgoulhon and J.-A. Marck, Phys. Rev. D {\bf 56};  
7740 (1997); H. Asada, Phys. Rev. D {\bf 57}, 7292 (1998).

\bibitem{sh98}
  M. Shibata, Phys. Rev. D {\bf 58}, 024012 (1998).  

\bibitem{te98}
  S. A. Teukolsky, Astrophys. J.  {\bf 504}, 442 (1998).  

\bibitem{EOS} For example, H. Heiselberg and M. Hjorth-Jensen, nucl-th/
9902033.

\bibitem{lrs93}
  D. Lai, F. A. Rasio and S. L. Shapiro, Astrophys. J. Suppl. 
{\bf 88}, 205 (1993); 
  D. Lai, F. A. Rasio and S. L. Shapiro, Astrophys. J. {\bf 420}, 811 (1994). 

\bibitem{lrs97}
  J. C. Lombardi, Jr., F. A. Rasio and S. L. Shapiro,
  Phys. Rev. D {\bf 56}, 3416 (1997).

\bibitem{st83} {\it e.g.}, 
S. L. Shapiro and S. A. Teukolsky,
  {\it Black Holes, White Dwarfs and Neutron Stars}, (Wiley,
  New York, 1983).

\bibitem{wm956} 
  J. R. Wilson and G. J. Mathews, Phys. Rev. Lett. {\bf 75}, 4161 (1995); 
  J. R. Wilson, G. J. Mathews and P. Marronetti, 
  Phys. Rev. D {\bf54}, 1317 (1996). 

\bibitem{bgm989}
S. Bonazzola, E. Gourgoulhon and J.-A. Marck,
to be published in the proceedings of 19th 
Texas Symposium on Relativistic Astrophysics: Texas in Paris, Paris, France 
(1998);
S. Bonazzola, E. Gourgoulhon and J.-A. Marck, 
Phys. Rev. Lett. {\bf 82}, 892 (1999).


\bibitem{mmr99}
 P. Marronetti, G. J. Mathews and J. R. Wilson, 
to be published in the proceedings of 19th
Texas Symposium on Relativistic Astrophysics: Texas in Paris, Paris, France
(1998);
  P. Marronetti, G. J. Mathews and J. R. Wilson, 
  Phys. Rev. D {\bf 60}, 087301 (1999). 


\bibitem{ue99}
  K. Ury\=u and Y. Eriguchi, Phys. Rev. D. {\bf 61}, 124023 (2000). 


\bibitem{foot3} 
{}From a post Newtonian point of view, the solution in 
the conformal flatness approximation is expected to deviate 
from the correct solution. The magnitude of the deviation is 
{\it e.g.}, of $O[(v/c))^4]$ for $\Omega$ and $J$ and 
of $O[(v/c))^6]$ for $M$ and $M_0$ where $v$ denotes 
characteristic magnitude of the orbital velocity: See, {\it e.g.}, 
H. Asada and M. Shibata, Phys. Rev. D {\bf 54}, 4944 (1996). 

\bibitem{uue99}
  F. Usui, K. Ury\=u and Y. Eriguchi, Phys. Rev. D {\bf 61}, 024039 (2000). 

\bibitem{ue98}
  K. Ury\=u and Y. Eriguchi, Mon. Not. R. astr. Soc. {\bf 296}, L1 (1998); 
Astrophys. J. Suppl. {\bf 118}, 563 (1998); 
Mon. Not. R. astr. Soc. {\bf 303}, 329 (1999); 
to be published in the proceedings of 19th
Texas Symposium on Relativistic Astrophysics: Texas in Paris, Paris, France
(1998).


\bibitem{commentcode} 
In the present computation, we typically took values of the 
type L mesh listed in Table II in \cite{ue99}.  
We also truncated the order of the Legendre expansion in Eq.~(69) 
and Eq.~(78) in \cite{ue99} at $n_{\rm max} = 32$ and 
$l_{\rm max} = 10$ or $12$, typically, instead of infinity.  
We have checked the convergence of the numerical scheme and 
the effect of the truncation due to the Legendre expansion 
in \cite{ue99}.  The above typical numbers for $n_{\rm max}$ and
$l_{\rm max}$ have been shown to give accurate results.  


\bibitem{commentsepa} 
We changed $\whd$ by steps of 0.125 typically to construct 
a solution sequence and by steps of 0.0625 near to 
the minima of $\bJ$ and $\bM$ when they appear on the sequence. 


\bibitem{footisco} 
Here, we refer to the general relativistic orbital instability as 
being that determined only by general relativistic gravity between 
two bodies and independent of any hydrodynamic effects and 
finite size effects, 
such as the orbital instability for a test particle 
orbiting near to a black hole. 
In this definition, the ISCO determined by the general relativistic 
instability should never depend on $\compa$ because of the strong 
equivalence principle. 




\bibitem{co94}
  G. B. Cook, Phys. Rev. D {\bf 50}, 5025 (1994).

\bibitem{B00}
T. W. Baumgarte, Phys. Rev. D (2000), to be published (gr-qc/0004050). 

\bibitem{kww92}
  L. E. Kidder, C. M. Will, A. G. Wiseman, Class. Quantum Gravity 9, 
  L125 (1992) ;  Phys. Rev. D {\bf 47}, 3281 (1993).  

\bibitem{dis98}
  T. Damour, B. R. Iyer, B. S. Sathyaprakash, 
  Phys. Rev. D {\bf 57}, 885 (1998). 

\bibitem{bd99}
  A. Buonanno, T. Damour, Phys. Rev. D {\bf 59}, 084006 (1999).


\bibitem{KIPeanna} 
E. E. Flanagan, Phys. Rev. D {\bf 58}, 124030 (1998): \\
K. S. Thorne, Phys. Rev. D {\bf 58}, 124031 (1998).


\bibitem{footq}

According to the quadrupole formula 
for gravitational radiation reaction, 
$J_{\rm tot}$ can decrease by $\sim 10\%$, 
but the fraction of decrease of $M_{\rm ADM}$ will be at most 
$\sim 1\%$ in one orbit around the innermost orbits. 
(In the terminology of the post Newtonian approximation, 
changes of $J_{\rm tot}$ and $M_{\rm ADM}$ are proportional to 
$(v/c)^5$ and $(v/c)^7$, respectively, where $v$ is a 
characteristic speed of the orbital velocity.) 
Thus, $J_{\rm tot}/M_{\rm ADM}^2$ is expected to decrease by $\sim 10\%$. 

\bibitem{BSS}
T. W. Baumgarte, S. L. Shapiro, and M. Shibata, Astrophys. J. 
{\bf 528}, L29 (2000); \\
M. Shibata, T. W. Baumgarte and S. L. Shapiro, Astrophys. J. 
to be published (astro-ph/0005378). 

\bibitem{ba97} 
  T. W. Baumgarte, G. B. Cook, M. A. Scheel, S. L. Shapiro and 
  S. A. Teukolsky, Phys. Rev. D {\bf 57}, 6181 (1998);
{\it ibid} {\bf 57}, 7299 (1998). 


\bibitem{footh3}
The hydrodynamic interaction could be an important mechanism for 
outward angular momentum transportation in particular 
for the case when spiral arms are formed in the outer region of a 
merger object and an ellipsoidal object are formed in the inner region.  
However, recent results of numerical simulations for the merger of 
irrotational binaries \cite{su00} show that such structures are not formed 
significantly, which indicates that the timescale of the transportation 
could not be short enough to increase the specific angular momentum of 
a certain fraction of fluid elements.  


\bibitem{footkb}
Note that $j$ at the ISCO for a test particle orbiting
progradely in the equatorial plane of a Kerr black hole
with $q=0.9$ is $j \approx 2.1M_{\rm BH}$ \cite{st83}.
(For the maximally rotating Kerr black hole $(q=1)$, $j$ at
the ISCO is $M_{\rm BH}$.)


\bibitem{ONS} {\it e.g.}, 
K. Oohara, T. Nakamura and M. Shibata, Prog. Theor. Phys. Suppl. 
{\bf 128}, 183 (1997). 

\bibitem{turning} R. Sorkin, Astrophys. J. {\bf 249}, 254 (1981); 
J. R. Ipser and G. Horowitz, Astrophys. J. {\bf 232}, 863 (1979); 
J. Katz, Mon. Not. R. astr. Soc. {\bf 183}, 765 (1978). 

\end{thebibliography}
\end{document}